\documentstyle[aps,psfig]{revtex}

\begin{document}


\title{The Nonlinear Meissner Effect in Unconventional Superconductors}
\author{D. Xu, S. K. Yip and J.A. Sauls}
\address{Department of Physics \& Astronomy,
Northwestern University, Evanston, IL 60208 USA}
\date{draft: July 17, 1994; revised: February 12, 1995}

\maketitle

\begin{abstract}

\centerline{{\bf Abstract}}
{\small
\noindent

We examine the long-wavelength current response in anisotropic
superconductors and show how the field-dependence of the Meissner
penetration length can be used to detect the structure of the order
parameter. Nodes in the excitation gap lead to a nonlinear
current-velocity constitutive equation at low temperatures which is
distinct for each symmetry class of the order parameter. The effective
Meissner penetration length is linear in $H$ and exhibits a
characteristic anisotropy for fields in the $ab$-plane that is
determined by the positions of the nodes in momentum space. The
nonlinear current-velocity relation also leads to an intrinsic magnetic
torque for in-plane fields that are not parallel to a nodal or
antinodal direction. The torque scales as $H^3$ for $T\rightarrow 0$
and has a characteristic angular dependence. We analyze the effects of
thermal excitations, impurity scattering and geometry on the current
response of a $d_{x^2-y^2}$ superconductor, and discuss our results in
light of recent measurements of the low-temperature penetration length
and in-plane magnetization of single-crystals of
$YBa_2Cu_3O_{7-\delta}$ and $LuBa_2Cu_3O_{7-\delta}$.
}
\end{abstract}



\bigskip


\section{Introduction}

Recent measurements of the Meissner penetration depth\cite{har93} and
Josephson interference effects in $YBa_2Cu_3O_{7-\delta}$\cite{wol93}
have been interpreted in support of a spin-singlet order parameter
belonging to the one-dimensional, $d_{x^2-y^2}$ representation,
$\Delta(\vec{p}_f)=\Delta_0\,(\hat{p}_x^2 - \hat{p}_y^2)$, which breaks
reflection symmetry in the basal plane. Such a pairing state has been
proposed by several authors\cite{bic87,mor90,mon91} based on arguments
that the CuO materials are Fermi liquids close to an SDW instability.

If the cuprates have an order parameter that is unconventional, {\it i.e.\ }
one
that breaks additional symmetries of the normal state besides gauge
symmetry, then the superconducting state is expected to exhibit a
number of novel properties, including (i) gapless excitations below
$T_c$, (ii) anomalous Josephson effects, (iii) exotic vortex structures
and associated excitations, (iv) new collective modes, (v) sensitivity
of superconducting coherence effects to defect scattering and (vi)
multiple superconducting phases.\cite{gor87,sig91} Many of these
signatures of unconventional pairing have been observed in superfluid
$^3$He, and in heavy fermion superconductors, notably
UPt$_3$.\cite{sau94} The case for an unconventional order parameter in
the cuprates, and particularly a $d_{x^2-y^2}$ state, is not settled;
there are conflicting interpretations of closely related
experiments,\cite{wol93,cha94,tsu94} variation in results that
are presumably related to material quality or
preparation,\cite{bon94,lee94} and experimental results that are not
easily accounted for within the $d_{x^2-y^2}$ model.\cite{sun94,bon93}

In this paper we examine the long-wavelength current response in
superconductors with an unconventional order parameter, and show how
the {\it field-dependence} of the Meissner penetration length can be
used to detect the structure of the order parameter. This report
extends our earlier work on nonlinear supercurrents,\cite{yip92a,xu94}
and provides the relevant analysis that could not be included in our
short reports. Specifically, we show (i) how the nodes in the
excitation gap, whose multiplicity and position in momentum space
depend on the symmetry class of the order parameter, lead to a
nonlinear current-velocity constitutive equation at low temperatures
($T\ll T_c$) which is unique and qualitatively distinct for each
symmetry class. The effective Meissner penetration length is {\it
linear} in $H$ and exhibits a characteristic anisotropy for fields in
the $ab$-plane. (ii) This anisotropy is determined by the {\it
positions} of the nodes in momentum space. For example, in the case of
a $d_{x^2-y^2}$ state in a tetragonal material the anisotropy is
precisely $1/\sqrt{2}$, independent of the detailed shape of the Fermi
surface or the gap. (iii) The nonlinear current-velocity relation leads
to an intrinsic magnetic torque for {\it in-plane} fields that are not
parallel to a nodal or antinodal direction. The torque scales as $H^3$
for $T\ll T_c$ and has a characteristic angular variation with a period
of $\pi/2$ (for tetragonal symmetry). The magnitude and angular
dependence of this torque are calculated for thick superconducting
films or slabs. (iv) We discuss the effects of thermal excitations,
impurity scattering and geometry for observing these features in a
$d_{x^2-y^2}$ superconductor. Recent measurements of the
low-temperature, zero-field penetration length\cite{har93} are used to
determine the relevant material parameters for $YBa_2Cu_3O_{7-\delta}$,
which are then used to estimate the magnitudes of the field-dependence
of the penetration depth and the torque anisotropy at low
temperatures.

Our starting point is Fermi-liquid theory applied to anisotropic
superconductors; section II includes the relevant theoretical framework
needed to calculate the current response in unconventional
superconductors. We derive formulas relating the equilibrium
supercurrent to the magnetic field and discuss the linear response
limit in section III. The nonlinear current-velocity constitutive
equation is examined in section IV. A clean superconductor with a line
of nodes in the gap has an anomalous contribution to the current which
is a nonanalytic function of the condensate velocity, $\vec{v}_s$, at
$T=0$. The relation of the anomalous current to the quasiparticle
spectrum is discussed, and the contribution of this current to the
Meissner penetration depth is obtained from solutions to the nonlinear
London equation. The effects of impurity scattering and thermally
excited quasiparticles on the anisotropy and field-dependence of the
supercurrent are examined in detail; the signatures of the anomalous
current survive thermal excitations and impurity scattering at
sufficiently low temperatures and weak (or dilute) impurity scattering.
We discuss our results in light of recent experiments on the
low-temperature penetration depth\cite{har93} in single crystals of
$YBa_2Cu_3O_{6.95}$. An important conclusion is that if the
linear temperature dependence of the penetration depth reported for
$YBa_2Cu_3O_{6.95}$ is due to the nodes of a $d_{x^2-y^2}$ order
parameter, then the nonlinear Meissner effect, including the intrinsic
anisotropy and field-dependence, should be observable for $T< 1\,K$
with a change in $\lambda_{ab}$ of approximately $30\,\AA$ over the field range
$0<H<H_{c1}\simeq 200\,G$. In section V we discuss the nonlinear
current, and associated in-plane magnetic torque, that develops for
surface fields that are not aligned along a nodal or antinodal
direction. The torque anisotropy (or transverse magnetization) is
obtained from solutions to the nonlinear London equation at low
temperatures. We also comment on a recent experimental report of a
measurement of the in-plane magnetization\cite{bua94} of a single
crystal of $LuBa_2Cu_3O_{7-\delta}$. In the rest of the introduction
we briefly discuss the symmetry classes and unconventional order
parameters for superconductors with tetragonal symmetry appropriate to
the $CuO$ superconductors (see Refs.(\onlinecite{vol84,sig91,yip93c})
for detailed discussions.)

\subsection*{Symmetries of the pairing state}

BCS superconductivity is based on a macroscopically occupied equal-time
pairing amplitude $f_{\alpha\beta}(\vec{p}_f)\sim
\left<a_{\vec{p}_f\alpha}\,a_{-\vec{p}_f\beta}\right>$, for
quasiparticle pairs near the Fermi surface with zero total momentum and
spin projections $\alpha$ and $\beta$. Fermi statistics requires that
the order parameter obey the anti-symmetry condition,
$f_{\alpha\beta}(\vec{p}_f)=-f_{\beta\alpha}(-\vec{p}_f)$, while
inversion symmetry (if present) implies that the pairing amplitude
decomposes into even-parity (spin-singlet) and odd-parity
(spin-triplet) sectors. Furthermore, the pairing interaction separates
into a sum over invariant bilinear products of basis functions for each
irreducible representation of the point group. The resulting
ground-state order parameter, barring the exceptional case of near
degeneracy in two different channels, belongs to a single irreducible
representation. For tetragonal symmetry there are four one-dimensional
(1D) representations and one two-dimensional (2D) representation, and
each of them occurs in both even- and odd-parity
representations.\footnote{The principal results and conclusions presented here
are
not qualitatively modified by a-b anisotropy; the quanitative effects
of a-b anisotropy will be discussed elsewhere.} The residual symmetry
of the order parameter is just that of the basis functions for the 1D
representations, but for the 2D representation there are three possible
ground states with different residual symmetry groups. There is no
evidence that we are aware of to support a spin-triplet order parameter
in the $CuO$ superconductors; in fact the temperature dependence of the
Knight shift in the cuprates\cite{tak89,bar90a} is argued to strongly
favor a spin-singlet order parameter.\cite{ann90} Thus, we limit the
discussion to even-parity, spin-singlet states; however, most of the
analysis and many of the main results for the current response are also
valid for odd-parity states.

\bigskip
\begin{table}
\caption{Even Parity Basis functions and Symmetry Classes for $D_{4h}$}
\label{tab_basis}
\begin{tabular}{clll}
Symmetry Class & Order Parameter: $\Delta(\vec{p}_f)$ & Residual Symmetry &
Nodes \\
\hline
$A_{1g}$ & 1 & $D_{4h}\times T$ & none \\

$A_{2g}$ &
$\hat{p}_x\hat{p}_y(\hat{p}_x^2-\hat{p}_y^2)$ &
$D_{4}[C_{4}]\times C_i \times T$ &
8 lines: $|\hat{p}_x|=\pm |\hat{p}_y|, \hat{p}_x =0, \hat{p}_y = 0$ \\

$B_{1g}$ &
$\hat{p}_x^2-\hat{p}_y^2 $ &
$D_{4}[D_{2}]\times C_i \times T$ &
4 lines: $|\hat{p}_x|=\pm |\hat{p}_y|$ \\

$B_{2g}$ &
$\hat{p}_x\hat{p}_y$ &
$D_{4}[D_{2}']\times C_i \times T$ &
4 lines: $\hat{p}_x =0, \hat{p}_y = 0 $ \\

$E_{g}(1,0)$ &
$\hat{p}_z\hat{p}_x$ &
$D_{2}[C_{2}']\times C_i \times T$ &
3 lines: $\hat{p}_z =0, \hat{p}_x = 0 $ \\

$E_{g}(1,1)$ &
$\hat{p}_z(\hat{p}_x +\hat{p}_y)$ &
$D_{2}[C_{2}'']\times C_i \times T$ &
3 lines: $\hat{p}_z =0, \hat{p}_x +\hat{p}_y = 0 $ \\

$E_{g}(1,i)$ &
$\hat{p}_z(\hat{p}_x +i\hat{p}_y)$ &
$D_{4}[E]\,\times C_i$ &
1 line: \,$\hat{p}_z =0$

\end{tabular}
\end{table}

Table I summarizes the symmetry classes of the order parameter for
spin-singlet pairing. All of the 1D representations have residual
symmetry groups which include four-fold rotations combined with
appropriate elements of the gauge groups. The states $E_g(1,0)$ and
$E_g(1,1)$ have a residual symmetry group that allows only two-fold
rotations. The resulting supercurrent, or superfluid density tensor,
for such states is in general strongly anisotropic in the basal plane.
The 2D order parameter, $E_g(1,i)$, preserves the four-fold rotational
symmetry, but breaks time-reversal symmetry.

Although the $B_{1g}$ ($d_{x^2-y^2}$) and $B_{2g}$ ($d_{xy}$) order
parameters break the $C_{4}$ rotational symmetry of the CuO planes, a
combined $C_{4}$ rotation and gauge transformation by $e^{i\pi}$ is a
symmetry. Since many properties of the superconducting state depend
only on Fermi-surface averages of $|\Delta(\vec{p}_f)|^2$, the broken
rotational symmetry is not easy to observe. In particular, the London
penetration depth tensor is cylindrically symmetric for any of the 1D
pairing states listed in Table I. Furthermore, all of the
unconventional gaps in Table I yield a linear temperature dependence at
$T\ll T_c$ for the zero-field penetration depth (in the clean limit).

A distinguishing feature of each phase, which is a consequence of their
particular broken symmetries, is that the nodes of each gap are located
in different positions in $\vec{p}$-space. A point that we make below
is that the field-dependence of the supercurrent may be used to locate
the positions of the nodal lines (or points) of an unconventional gap
in momentum space. This gap spectroscopy is possible at low
temperatures, $T\ll T_c$, and is based on features which are intrinsic
to nearly all unconventional BCS states in tetragonal or orthorhombic
structures.

\section{Fermi-liquid theory of superconductivity}

Our starting point for calculations of the current response is the
Fermi-liquid theory of superconductivity. This theory is general enough
to include real materials effects of Fermi-surface anisotropy, impurity
scattering and inelastic scattering from phonons and quasiparticles, in
addition to unconventional pairing.  A basic feature of the
Fermi-liquid theory of superconductivity ({\it c.f.}
Refs.(\onlinecite{ser83,rai86,rai94}) for a more detailed discussion of
the formulation of Fermi-liquid theory.) is that for low excitation
energies $(\hbar\omega, k_BT, \hbar qv_f, \Delta)\ll E_f$, the wave
nature of the quasiparticle excitations is unimportant and can be
eliminated by integrating the full Matsubara Green's function over the
quasiparticle momentum (or kinetic energy) in the low-energy band
around the Fermi surface,
\begin{equation}\label{QCg}
g_{\alpha\beta}(\vec{p}_f,\vec{R};\epsilon_n)=-
\int^{\omega_c}_{-\omega_c}\,d\xi_{\vec{p}}\,
\int_0^{\beta}\,d\tau\,e^{i\epsilon_n\tau/\hbar}
\int\,d\vec{r}\,e^{-i\vec{p}\cdot\vec{r}/\hbar}
\,<T_{\tau}\,\psi_{\alpha}(\vec{R}+\vec{r}/2,\tau)\,
\psi^{\dagger}_{\beta}(\vec{R}-\vec{r}/2,0)>\,,
\end{equation}
where $\xi_{\vec{p}}=v_f(\vec{p}_f)(|\vec{p}|-|\vec{p}_f|)$ is the
normal-state quasiparticle excitation energy for momentum $\vec{p}$
nearest to the position $\vec{p}_f$ on the Fermi surface and
$\vec{v}_f(\vec{p}_f)$ is the quasiparticle velocity at the point
$\vec{p}_f$. The resulting quasiclassical propogator is a function of
the momentum direction $\vec{p}_f$ on the Fermi surface, the center of
mass coordinate $\vec{R}$ and the Matsubara energy
$\epsilon_n=(2n+1)\pi T$. The pairing correlations are described by the
$\xi$-integrated anomalous Green's functions,
\begin{equation}
f_{\alpha\beta}(\vec{p}_f,\vec{R};\epsilon_n)=-
\int^{\omega_c}_{-\omega_c}\,d\xi_{\vec{p}}\,
\int_0^{\beta}\,d\tau\,e^{i\epsilon_n\tau/\hbar}
\int\,d\vec{r}\,e^{-i\vec{p}\cdot\vec{r}/\hbar}
\,<T_{\tau}\,\psi_{\alpha}(\vec{R}+\vec{r}/2,\tau)\,
\psi_{\beta}(\vec{R}-\vec{r}/2,0)>\,.
\end{equation}
The low-energy quasiparticle spectrum, combined with charge
conservation and gauge invariance, allows one to formulate observables
in terms of the quasiclassical Green's function and material parameters
defined on the Fermi surface. For example, the equilibrium current is
given by
\begin{equation}\label{QCcurrent}
\vec{j}_s(\vec R)=-eN_f\int d\vec{p}_f\,\vec{v}_f(\vec{p}_f)\, T\sum_n
\frac{1}{2}\,Tr\left\{\hat{\tau}_3\,\hat{g}(\vec{p}_f, \vec R;
\epsilon_n)\right\}\,,
\end{equation}
where $N_f$ is the single-spin density of states at the Fermi level and
the integration is over the Fermi surface with a weight factor of the
angle-resolved density of states normalized to
unity. We have introduced the $4\times 4$ quasiclassical functions in
`spin $\times$ particle-hole' space; a convenient
representation for the particle-hole and spin structure of the
propagator is,
\begin{equation}\label{QCmatrix}
\hat{g}=
\left(
\begin{array}{cc}
g(\vec{p}_f,\vec{R};\epsilon_n) +
\vec{g}(\vec{p}_f,\vec{R};\epsilon_n)\cdot\vec{\sigma} &
f(\vec{p}_f,\vec{R};\epsilon_n)\,i\sigma_y +
\vec{f}(\vec{p}_f,\vec{R};\epsilon_n)\cdot
			   i\vec{\sigma}\sigma_y \\
f(-\vec{p}_f,\vec{R};\epsilon_n)^*\,i\sigma_y -
                           \vec{f}(-\vec{p}_f,\vec{R};\epsilon_n)^*\cdot
			   i\sigma_y\vec{\sigma}\quad &
\quad g(-\vec{p}_f,\vec{R};-\epsilon_n) -
\vec{g}(-\vec{p}_f,\vec{R};-\epsilon_n)\cdot\sigma_y\vec{\sigma}\sigma_y
\end{array}
\right)
\,.
\end{equation}
This matrix structure represents the remaining quantum mechanical
degrees of freedom; the coherence of particle and hole
states is
contained in the off-diagonal elements in eq.(\ref{QCmatrix}). The
diagonal components are separated into spin-scalar, $g$, and
spin-vector, $\vec{g}$, components. The scalar component determines the
current response, while the vector components determine the
spin-paramagnetic response. The off-diagonal propagator separates into
spin-singlet, $f$, and spin-triplet, $\vec{f}$, pairing amplitudes,
which are coupled to the diagonal propagators through the
quasiclassical transport equation,
\begin{equation}\label{trans}
\hat{Q}[\hat{g}\,,\,\hat{\sigma}]\equiv
\left[i\epsilon_n\hat{\tau}_3-\hat{\sigma}(\vec{p}_f, \vec{R};\epsilon_n)\,,\,
\hat{g}(\vec{p}_f, \vec{R};\epsilon_n)\right]
+i\vec{v}_f\cdot\vec{\nabla}\hat{g}(\vec{p}_f, \vec{R};\epsilon_n)=0
\,,
\end{equation}
first derived by Eilenberger\cite{eil68} by eliminating the
high-energy, short-distance structure of the full Green's function in
Gorkov's equations.\cite{gor59} The transport equation is supplemented
by the normalization condition,
\begin{equation}\label{norm}
\hat{g}(\vec{p}_f,\vec{R};\epsilon_n)^2 = -\pi^2\,\hat{1}
\,,
\end{equation}
which eliminates many unphysical solutions from the general set of solutions to
the transport equation.\cite{eil68}

The self energy, $\hat{\sigma}$, has an expansion (Fig. 1) in terms of
$\hat{g}$ (solid lines) and renormalized vertices describing the
interactions between quasiparticles, phonons (wiggly lines), impurities
and external fields. An essential feature of Fermi liquid theory is
that this expansion is based on a set of small expansion parameters,
${\sl small}\sim{k_BT_c}/{E_f}\,,{\hbar}/{p_f\xi_0}\,,\,...\ll 1$, which are
the relevant low-energy ({\it e.g.\ } pairing energy) or
long-wavelength ({\it e.g.\ } coherence length) scales compared to the
characteristic high-energy ({\it e.g.\ } Fermi energy) or short-wavelength
({\it e.g.\ }
Fermi wavelength) scales.\cite{ser83} The leading order contributions
to the self energy are represented in Fig.1. Diagram (1a) is the
zeroth-order in ${\sl small}$ and represents the band-structure
potential of the quasiparticles. This term is included as Fermi-surface
data for $\vec{p}_f$, $\vec{v}_f$ and $N_f$, which is taken from
experiment or defined by a model for the band-structure. Diagram (1b)
is first-order in ${\sl small}$ and represents Landau's Fermi-liquid
interactions (diagonal in particle-hole space), and the electronic
pairing interactions (off-diagonal in particle-hole space), or
mean-field pairing self-energy (also the `order parameter' or `gap
function'). Diagram (1b') represents the leading-order phonon contribution to
the electronic self energy (diagonal) and pairing self-energy (off-diagonal);
however, we confine our discussion to electronically driven superconductivity
with a frequency-independent interaction.

\begin{figure}
\centerline{\psfig{figure=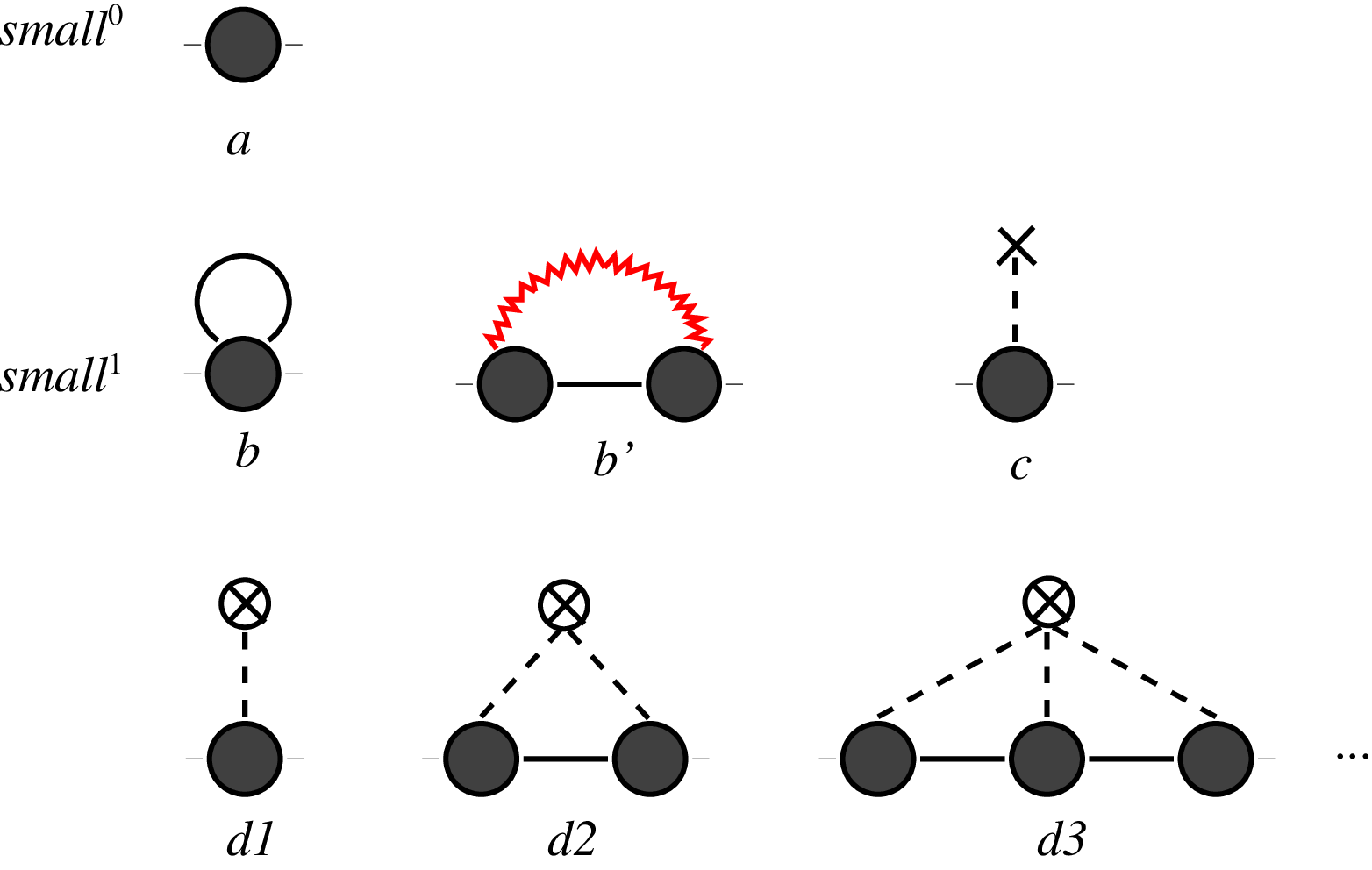,height=3in}}
\begin{quote}
\small
Fig. 1{\hskip 10pt}
Leading order contributions to the quasiclassical self-energy.
\end{quote}
\end{figure}

In the spin-singlet channel, the order parameter satisfies the gap
equation,
\begin{equation}
\Delta(\vec{p}_f,\vec{R})=\int d\vec{p}\,'\!\!\!_f\
V(\vec{p}_f,\vec{p}\,'\!\!\!_f)\ T\sum_{\epsilon_n}
f(\vec{p}\,'\!\!\!_f,\vec{R};\epsilon_n)
\,,
\end{equation}
where $f(\vec{p}_f,\vec{R};\epsilon_n)$ is the spin-singlet pairing
amplitude and $V(\vec{p}_f,\vec{p}\,'\!\!\!_f)$ represents the
electronic pairing interaction; this function may be expanded in basis
functions for the irreducible representations of the point group,
\begin{equation}
V(\vec{p}_f,\vec{p}\,'\!\!\!_f)=\sum^{irrep}_{\alpha}
V_{\alpha}\,\sum^{d_{\alpha}}_{i=1}{\cal Y}_{\alpha i}(\vec{p}_f){\cal
Y}_{\alpha i}(\vec{p}\,'\!\!\!_f)\,,
\end{equation}
where the parameter, $V_{\alpha}$, is the pairing interaction in the channel
labeled by the $\alpha$th irreducible representation, and the corresponding
basis functions, $\{{\cal Y}_{\alpha i}(\vec{p}_f)|i=1,...,d_{\alpha}\}$, are
orthonormal,
$\left<{\cal Y}_{\alpha i}(\vec{p}_f)
{\cal Y}_{\beta j}^*(\vec{p}_f)\right>_{\vec{p}_f}
=\delta_{\alpha\beta}\delta_{ij}$,
where the Fermi surface average is defined by
$<A(\vec{p}_f)>_{\vec{p}_f}=\int
d\vec{p}_f\ A(\vec{p}_f)$.

\subsection*{Impurity scattering}

The summation of diagrams (1d) gives the leading-order self energy from a
random distribution of impurities in terms of the impurity
t-matrix,\cite{buc79,buc81}
\begin{equation}
\hat{\sigma}_{imp}(\vec{p}_f;\epsilon_n)=
n_{imp}\,\hat{t}(\vec{p}_f,\vec{p}_f;\epsilon_n)
\,,
\end{equation}
\begin{equation}
\hat{t}(\vec{p}_f,\vec{p}\,'\!\!\!_f;\epsilon_n)=
\hat{u}(\vec{p}_f,\vec{p}\,'\!\!\!_f)+
N_f\int d\vec{p}\,''\!\!\!_f\, \hat{u}(\vec{p}_f,\vec{p}\,''\!\!\!_f)\,
\hat{g}(\vec{p}\,''\!\!\!_f;\epsilon_n)
\,\hat{t}(\vec{p}\,''\!\!\!_f,\vec{p}\,'\!\!\!_f;\epsilon_n)
\,.
\end{equation}
The first term is the matrix element of the impurity potential between
quasiparticles at points $\vec{p}_f$ and $\vec{p}\,'\!\!\!_f$ on the
Fermi surface, $n_{imp}$ is the impurity concentration, and the
intermediate states are defined by the self-consistently determined
quasiclassical propagator.

For a spin-singlet superconductor with non-magnetic
impurities, $\hat{u}(\vec{p}_f,\vec{p}\,'\!\!\!_f)
=u(\vec{p}_f,\vec{p}\,'\!\!\!_f)\,\hat{1}$, and the terms in
$\hat{\sigma}_{imp}$ that contribute in the
transport equation lead to a renormalization of the
Matsubara frequency and gap function;
$i\tilde{\epsilon}_n=
i\epsilon_n-\sigma_{imp}(\vec{p}_f;\epsilon_n)$
and
$\tilde{\Delta}(\vec{p}_f;\epsilon_n)=
\Delta(\vec{p}_f)+\Delta_{imp}(\vec{p}_f;\epsilon_n)$.
Thus, the solution to the transport equation and normalization
condition for the propagator becomes,
\begin{equation}\label{g0}
\hat{g}(\vec{p}_f;\epsilon_n) =
-\pi\frac{i\tilde{\epsilon}_n(\vec{p}_f;\epsilon_n)\hat{\tau}_3 -
\hat{\tilde{\Delta}}(\vec{p}_f;\epsilon_n)}
{\sqrt{\tilde{\epsilon}_n(\vec{p}_f;\epsilon_n)^2 +
|\tilde{\Delta}(\vec{p}_f;\epsilon_n)|^2}}
\,.
\end{equation}

In the second-order Born approximation for the impurity t-matrix
(this is not essential, but simplifies the following discussion). The
impurity renormalization of the off-diagonal self-energy is given by,
\begin{equation}\label{renormdelta}
\tilde{\Delta}(\vec{p}_f;\epsilon_n) = \Delta(\vec{p}_f)+
\int\,d{\vec{p}\,'\!\!\!_f}\,
w(\vec{p}_f,\vec{p}\,'\!\!\!_f)
\frac{\tilde{\Delta}(\vec{p}\,'\!\!\!_f;\epsilon_n)}
{\sqrt{\tilde{\epsilon}_n^2+|\tilde{\Delta}(\vec{p}\,'\!\!\!_f;\epsilon_n)|^2}}
\,,
\end{equation}
where $w(\vec{p}_f,\vec{p}\,'\!\!\!_f)=2\pi n_{imp}\,N_f
|u(\vec{p}_f,\vec{p}\,'\!\!\!_f)|^2$ is the scattering rate in the Born
approximation. Note that the integral equation for
$\tilde{\Delta}(\vec{p}_f;\epsilon_n)$ has the mean-field order parameter,
$\Delta(\vec{p}_f)$, as the driving term. The scattering rate
$w(\vec{p}_f,\vec{p}\,'\!\!\!_f)$ has the full symmetry of the normal metal;
thus, it too can be expanded in basis functions for the irreducible
representations of the point group,
\begin{equation}
w(\vec{p}_f,\vec{p}\,'\!\!\!_f)=
\sum_{\alpha}^{irrep}\,\frac{1}{2\tau_{\alpha}}\,
\sum_{i=1}^{d_{\alpha}}\,{\cal Y}_{\alpha i}(\vec{p}_f)
{\cal Y}^*_{\alpha i}(\vec{p}\,'\!\!\!_f)
\,,
\end{equation}
where $1/2\tau_{\alpha}$ is the scattering rate for channel $\alpha$.
The integral equation for the renormalized order parameter separates
into algebraic equations for each representation,
\begin{equation}
\tilde{\Delta}_{\alpha i} =
\int\,d\vec{p}_f\,{\cal Y}^*_{\alpha i}(\vec{p}_f)\,
\Delta(\vec{p}_f) + \frac{1}{2\tau_{\alpha}}
\int\,d\vec{p}_f\,
\frac{{\cal Y}^*_{\alpha i}(\vec{p}_f)\,
\tilde{\Delta}(\vec{p}_f;\epsilon_n)}
{\sqrt{\tilde{\epsilon}_n^2+|\tilde{\Delta}(\vec{p}_f;\epsilon_n)|^2}}
\,.
\end{equation}
The driving term is non-zero only for the irreducible representation
corresponding to $\Delta(\vec{p}_f)$. Thus, the resulting solution for
the impurity renormalized order parameter necessarily has the same
orbital symmetry as the mean-field order parameter; and the magnitude
of the impurity renormalization is determined by the scattering
probability for scattering in the same channel as that of
$\Delta(\vec{p}_f)$.\cite{gor85} The argument also holds for the full t-matrix.

For isotropic (`s-wave') impurity scattering the renormalized Matsubara
frequency and order parameter become
\begin{equation}
\tilde{\epsilon}_n=\epsilon_n +\frac{1}{2\tau}
\left<
\frac{\tilde{\epsilon}_n}
{\sqrt{\tilde{\epsilon}_n^2+
|\tilde{\Delta}(\vec{p}\,'\!\!\!_f;\epsilon_n)|^2}}
\right>_{\vec{p}\,'\!\!\!_f}
\,,
\end{equation}
\begin{equation}\label{sgaprenorm}
\tilde{\Delta}(\vec{p}_f;\epsilon_n) = \Delta(\vec{p}_f)+\frac{1}{2\tau}
\left<
\frac{\tilde{\Delta}(\vec{p}\,'\!\!\!_f;\epsilon_n)}
{\sqrt{\tilde{\epsilon}_n^2+|\tilde{\Delta}(\vec{p}\,'\!\!\!_f;\epsilon_n)|^2}}
\right>_{\vec{p}\,'\!\!\!_f}
\,.
\end{equation}
Thus, for an s-wave order parameter
these equations give identical renormalization factors for both
the Matsubara frequency and the order parameter, {\it i.e.\ }
\begin{equation}
{\tilde{\epsilon}_n\over\epsilon_n}={\tilde{\Delta}(\epsilon_n)\over\Delta}=
Z(\epsilon_n)=1+\frac{1}{2\pi\tau}\frac{1}{\sqrt{\epsilon_n^2 + \Delta^2}}
\qquad (s-wave) \,,
\end{equation}
in which case the impurity renormalization drops out of the equilibrium
propagator and gap equation.\cite{and59,abr59b} However,
s-wave superconductors are exceptional; for any unconventional order
parameter impurity scattering is pairbreaking.\cite{buc81}

Consider an unconventional superconductor with impurities in which the
scattering is dominated by the identity representation. If there is an
element of the point group, ${\cal R}$, which changes the sign of
$\Delta(\vec{p}_f)$, {\it i.e.\ } $\Delta(\vec{p}_f){\buildrel{\cal
R}\over\longrightarrow} - \Delta(\vec{p}_f)$, then from
eq.(\ref{sgaprenorm}) the impurity renormalization of the order
parameter vanishes identically:
$\tilde{\Delta}(\vec{p}_f;\epsilon_n)=\Delta(\vec{p}_f)$. The
cancellation between the impurity renormalization factors for the
Matsubara frequency and order parameter no longer occurs, with the
consequence that impurity scattering suppresses both $T_c$ and the
magnitude of the order parameter.

For isotropic impurity scattering (not restricted to the Born
approximation), the renormalization factor for the Matsubara frequency,
$\tilde{\epsilon}_n /\epsilon_n= Z(\epsilon_n)$, is independent of
position on the Fermi surface and given by
\begin{equation}\label{Z_imp}
Z(\epsilon_n) = 1 + \Gamma_u
\frac{Z(\epsilon_n)\,{\cal D}(\epsilon_n)}
{{\rm ctn}^2(\delta_0) + (Z(\epsilon_n)\,\epsilon_n\,{\cal D}(\epsilon_n))^2}
\,,
\end{equation}
with
\begin{equation}
{\cal D}(\epsilon_n) =
\left<
\frac{1}{\sqrt{Z(\epsilon_n)^2\epsilon_n^2+|\Delta(\vec{p}_f)|^2}}
\right>_{\vec{p}_f}
\,,
\end{equation}
where $\Gamma_u=n_{imp}/\pi N_f$ and $\delta_0=\tan^{-1}(\pi N_f u_0)$ is
the s-wave scattering phase shift in the normal state. In the Born
limit, $\delta_0\rightarrow \pi N_f u_0$, while in the strong
scattering limit ($N_f u_0\rightarrow\infty$) we obtain the unitarity
limit, $\delta_0\rightarrow\pi/2$.

Given the gap function, $\Delta(\vec{p}_f)$, the impurity
renormalization is easily calculated. The magnitude and temperature
dependence of the order parameter are calculated
self-consistently from the mean-field gap equation,
\begin{equation}\label{BCSgap}
\Delta(\vec{p}_f)=
\int\,d\vec{p}\,'\!\!\!_f
\,V(\vec{p}_f,\vec{p}\,'\!\!\!_f)\,
\pi T\sum_{\epsilon_n}^{|\epsilon_n|<\omega_c}\,
\frac{\Delta(\vec{p}\,'\!\!\!_f)}{\sqrt{Z(\epsilon_n)^2\epsilon_n^2+
|\Delta(\vec{p}\,'\!\!\!_f)|^2}}
\,.
\end{equation}
The linearized gap equation determines $T_c$ in terms of the pairing
interaction, frequency cutoff $\omega_c$ and impurity scattering rate.
At $T_c$ only the dominant pairing channel $\alpha$ is relevant and the
linearized gap equation becomes,
\begin{equation}
\frac{1}{V_{\alpha}}=\pi T_c \sum_{\epsilon_n}^{|\epsilon_n| < \omega_c}
\frac{1}{|\epsilon_n|+\Gamma},
\end{equation}
where $\Gamma=\Gamma_u\sin^2\delta_0$ is the pair-breaking parameter,
\begin{equation}\label{gamma}
\Gamma=\left\{
\begin{array}{l}
\Gamma_u\sin^2\delta_0=\frac{1}{2\tau}=\pi n_{imp}N_f u_0^2\quad\quad \quad
,\qquad ({\rm Born \,\,limit}) \\
\Gamma_u=\frac{n_{imp}}{\pi N_f}\quad \qquad \qquad\qquad\qquad \qquad
,\qquad({\rm unitarity\,\,limit})\,.
\end{array}
\right.
\end{equation}
For $\Gamma=0$, this equation determines the clean-limit value of the
transition temperature, $T_{co}$. Eliminating the pairing interaction
and cutoff gives the well-known Abrikosov-Gorkov formula,\cite{abr61}
except that the pairbreaking parameter is determined by non-magnetic
scattering,\cite{gor87}
\begin{equation}\label{Tc_reduction}
\psi\left(\frac{1}{2}+\frac{\Gamma}{2\pi T_c}\right) -
\psi\left(\frac{1}{2}\right)= \ln\left(\frac{T_{co}}{T_c}\right)
\,,
\end{equation}
where $\psi(z)$ is the digamma function.

Finally, the linearized gap equation is used to eliminate the pairing
interaction and cutoff in favor of $T_c$ in the full gap equation
(\ref{BCSgap}). For pairing in a 1D representation,
$V(\vec{p}_f,\vec{p}\,'\!\!\!_f)=
V\,e(\vec{p}_f)e(\vec{p}\,'\!\!\!_f)^*$, the order parameter is
$\Delta(\vec{p}_f)=\Delta\,e(\vec{p}_f)$, where $\Delta$ is obtained
from eq.(\ref{BCSgap}). Multiplying eq.(\ref{BCSgap}) by
$e(\vec{p}_f)^*$, integrating over the Fermi surface, and adding and
subtracting the RHS of the linearized gap equation (with
$T_c\rightarrow T$) to eliminate $V$ gives,
\begin{eqnarray}\label{gap(T)}
\lefteqn{\left[\ln(T/T_c)+\psi(\frac{1}{2}+\frac{\Gamma}{2\pi T})
-\psi(\frac{1}{2}+\frac{\Gamma}{2\pi T_c})\right] = } \nonumber \\
& & 2\pi T\sum_{n=0}^{\infty}
\left[
\left<\frac{|e(\vec{p}_f)|^2}{\sqrt{Z(\epsilon_n)^2\epsilon_n^2 +
\Delta^2 |e(\vec{p}_f)|^2}}
\right>_{\vec{p}_f}
-\frac{1}{\epsilon_n +\Gamma }
\right]\,,
\end{eqnarray}
which is solved self-consistently with eq.(\ref{Z_imp}) to give
$Z(\epsilon_n)$ and $\Delta$ as a function of $T/T_c$ and $\Gamma$.

\subsection*{Gauge-invariant coupling to the condensate flow field}

The self-energy term representing the diamagnetic coupling of
quasiparticles to a static magnetic field (Fig. 1c) is determined by
local gauge invariance. Under a gauge transformation of the Fermion
fields,
$\psi(\vec{r})\rightarrow\psi(\vec{r})e^{-i\Lambda(\vec{r})/2}$, the
quasiclassical propagator transforms as
\begin{equation}
\hat{g}{\buildrel\Lambda\over\longrightarrow}
\hat{g}'=\hat{U}(\Lambda)^{\dagger}
\ \hat{g}\ \hat{U}(\Lambda)
\,,
\end{equation}
where
$\hat{U}(\Lambda)=\exp(+\frac{i}{2}\Lambda(\vec{R})\hat{\tau}_3)$, as
does the self-energy and order parameter.\cite{ser83} Applying this
transformation
to the transport equation (\ref{trans}) gives
$\hat{U}(\Lambda)^{\dagger}\ \hat{Q}[\hat{g},\hat{\sigma}]
\ \hat{U}(\Lambda)
=\hat{Q}[\hat{g}',\hat{\sigma}'+\hat{\sigma}_{\nabla}]$. Thus, the
form of the transport equation is invariant, but the local gauge field
generates an additional self-energy,
$\hat{\sigma}_{\nabla}=-i\hat{U}(\Lambda)^{\dagger}
\ \vec{v}_f(\vec{p}_f)\cdot\vec{\nabla}\ \hat{U}(\Lambda)$. This
property of the transport equation is used to eliminate the phase
degree of freedom of the order parameter in favor of a spatially
varying flow field. We parametrize the spatial variations in terms of a
physical gauge, the phase $\chi(\vec{R})$, and the local amplitude,
$\Delta_0(\vec{p}_f;\vec{R})=|\Delta(\vec{R})|\,e(\vec{p}_f)$
\begin{equation}
\hat{\Delta}(\vec{p}_f,\vec{R})=
\hat{U}[\chi(\vec{R})]
\,\hat{\Delta}_0(\vec{p}_f,\vec{R})\,\hat{U}^{\dagger}[\chi(\vec{R})]
\,.
\end{equation}
Thus, the transport equation becomes
\begin{equation}\label{QCtrans2}
\left[i\epsilon_n\hat{\tau}_3 -\hat{\Delta}_0 -\hat{\sigma}_{v}
- \hat{\sigma}^{'}\,,\,\hat{g}'\right]
+i\vec{v}_f(\vec{p}_f)\cdot \vec{\nabla}\hat{g}'=0
\,,
\end{equation}
where
$\hat{\sigma}_{v}=\frac{1}{2}\vec{v}_f(\vec{p}_f)\cdot\vec{\nabla}\chi\,\hat{\tau}_3$. The diamagnetic coupling to a magnetic field, $\vec{b}=\vec{\nabla}\times\vec{A}$, is then determined
by gauge invariance, and can be represented in terms of the gauge-invariant
condensate flow field,
\begin{equation}
\vec{v}_s = {1\over 2}(\vec{\nabla}\chi
+ {2e\over c}\,\vec{A})
\,,
\end{equation}
and the self-energy,
\begin{equation}
\hat{\sigma}_{v}=\vec{v}_f(\vec{p}_f)\cdot\vec{v}_s(\vec{R})\,\hat{\tau}_3
\,.
\end{equation}

In the Meissner geometry ($\vec{H}$ parallel to the interface) the
driving term associated with the applied surface field is of order,
\begin{equation}
\left|\frac{\sigma_v}{\pi T_c}\right|
\sim \frac{\frac{e}{c}v_f A}{\pi T_c}
\sim\frac{H}{H_c}
\,,
\end{equation}
where $H_c\sim\phi_0/(\xi\lambda)$ is the thermodynamic critical field.
This term should be compared to the gradient term arising from spatial
variations of the screening current,
\begin{equation}
\left|\frac{\vec{v}_f\cdot\vec{\nabla}g}{\pi T_c}\right|
\sim\left|\frac{\vec{v}_f\cdot\vec{\nabla}(\sigma_v/\Delta)}{\pi T_c}\right|
\sim\frac{\xi}{\lambda}\frac{H}{H_c}
\,.
\end{equation}
In the strong type II limit the velocity field is effectively uniform
on the scale of the coherence length so we are generally justified in
dropping the gradient term in eq. (\ref{QCtrans2}).

\subsection*{Linear response}

The linear response limit is simply obtained from a perturbation
expansion of the propagator, transport equation and normalization
condition, and is expected to be valid for low magnetic fields,
$\left|{\sigma_v}/{\pi T_c}\right| \sim{H}/{H_c} \ll 1$.
Assume an expansion of the form, $\hat{g}=\hat{g}_0+\hat{g}_1+ ...$,
where $\hat{g}_0$ is the zero-field solution to the transport equation
given by eq.(\ref{g0}), and $\hat{g}_1$ is the first-order correction
to the propagator, formally of order $|\hat{g}_1|\sim {\cal
O}|(\sigma_v/\Delta)\hat{g}_0|$. The linearized transport equation and
normalization condition,
\begin{equation}
\left[
i\tilde{\epsilon}_n\hat{\tau}_3 - \hat{\tilde{\Delta}}
\,,
\hat{g}_1
\right]
-\left[\hat{\sigma}_v\,,\hat{g}_0\right]=0
\qquad\,,\qquad
\left\{\hat{g}_0\,,\hat{g}_1\right\}=0
\,,
\end{equation}
are inverted with the aid of eqs. (\ref{g0}) and (\ref{norm}) to give
\begin{equation}\label{g1}
\hat{g}_1=\pi\hat{\sigma}_{v}
{|\tilde{\Delta}|^2-i\tilde{\epsilon}_n\hat{\tau}_3\hat{\tilde{\Delta}}
\over{[\tilde{\epsilon}_n^2+|\tilde{\Delta}(\vec{p}_f,\epsilon_n)|^2]^{3/2}}}
\,.
\end{equation}
The resulting supercurrent calculated from
eq. (\ref{QCcurrent}) can be written in terms of the superfluid
density tensor,
\begin{equation}\label{rhos}
\left({\buildrel\leftrightarrow\over\rho}_s\right)_{ij} =
2N_f\int\,d\vec{p}_f\,\Phi(\vec{p}_f)\,v_f^i(\vec{p}_f)\,v_f^j(\vec{p}_f)
\,,
\end{equation}
where
\begin{equation}\label{Yosida}
\Phi(\vec{p}_f)=\pi T\sum_{\epsilon_n}
{|\tilde{\Delta}(\vec{p}_f,\epsilon_n)|^2\over
[\tilde{\epsilon}_n^2+|\tilde{\Delta}(\vec{p}_f,\epsilon_n)|^2]^{3/2}}
\,,
\end{equation}
which reduces to the angle-dependent Yosida function in the clean
limit.

\subsection*{Fermi-liquid effects}

Fermi-liquid effects arise from the leading order electronic
self-energy (diagram 1b). For the diamagnetic response the most
important Fermi-liquid effect is the  screening correction to the
diamagnetic current; the relevant self energy is,
$\hat{\sigma}_{b}(\vec{p}_f,\vec{R})=
\sigma_{b}(\vec{p}_f,\vec{R})\,\hat{\tau}_3$,
with
\begin{equation}\label{sigma_fl}
\sigma_{b}(\vec{p}_f,\vec{R})=
\int\,d\vec{p}\,'\!\!\!_f\,
A^{cur}(\vec{p}_f,\vec{p}\,'\!\!\!_f)\,T\sum_{\epsilon_n}\,g(\vec{p}\,'\!\!\!_f,\vec{R};\epsilon_n)
\,,
\end{equation}
where $A^{cur}(\vec{p}_f,\vec{p}\,'\!\!\!_f)$ is the dimensionless
quasiparticle interaction.

Fermi liquid effects can contribute substantial temperature-dependent
corrections to the penetration depth as the gap opens and the number of
thermal quasiparticles drops rapidly. We include formulas for the
Fermi-liquid correction to the supercurrent for a model uniaxial Fermi
surface. The position on the Fermi surface can be parametrized by the
direction of the Fermi wavevector, $\hat{p}$, and the Fermi velocity is
given by,
\begin{equation}
\vec{v}_f=v_f^{||} (\hat{p}_x \hat{x} + \hat{p}_y \hat{y})
+ v_f^{\perp}\hat{p}_z\hat{z}
\,.
\end{equation}
Similarly, the quasiparticle interaction is parametrized by
two Landau parameters corresponding to current flow in the basal plane
and along the $\hat{z}$-axis,
\begin{equation}\label{FL_cur}
A^{cur}(\vec{p}_f,\vec{p}\,'\!\!\!_f)=
A^{||}\left(\hat{p}_x\hat{p}'_x + \hat{p}_y\hat{p}'_y\right)
+
A^{\perp}\left(\hat{p}_z\hat{p}'_z\right)
\,.
\end{equation}
The linear response result for the supercurrent is easily
obtained from eq.(\ref{g1}) with the replacement
$\hat{\sigma}_v\rightarrow\hat{\sigma}_{sc}=
\vec{v}_f(\vec{p}_f)\cdot\vec{v}_s+\sigma_{b}(\vec{p}_f)$. The resulting
current is given by
\begin{equation}
\vec{j}_s(\vec{R})=-2eN_f\int d\vec{p}_f\,
\vec{v}_f(\vec{p}_f)\,\Phi(\vec{p}_f)\,\sigma_{sc}(\vec{p}_f)
\,,
\end{equation}
with the screening field satisfying the self-consistency equation,
\begin{equation}\label{molfield}
\sigma_{sc}(\vec{p}_f)= \vec{v}_f(\vec{p}_f)\cdot\vec{v}_s +
\int d\vec{p}\,'\!\!\!_f\, A^{cur}(\vec{p}_f,\vec{p}\,'\!\!\!_f)\,
\Phi(\vec{p}\,'\!\!\!_f)\,\sigma_{sc}(\vec{p}\,'\!\!\!_f)
\,.
\end{equation}
Equations
(\ref{rhos},\ref{Yosida},\ref{FL_cur},\ref{molfield},\ref{Z_imp},\ref{gap(T)})
are the basic equations used to calculate the temperature-dependent
penetration depth for unconventional superconductors.

For a gap $|\Delta(\vec{p}_f)|$ with four-fold rotational symmetry about the
$\hat{z}$-axis the resulting supercurrent is given by a diagonal superfluid
density tensor with in-plane ($\rho_s^{||}$) and z-axis ($\rho_s^{\perp}$)
superfluid densities given by
\begin{equation}\label{rhos_fl}
\rho_s^{||,\perp} = 2N_f(v_f^{||,\perp})^2
\frac{\phi^{||,\perp}}{1-\frac{1}{3}A^{||,\perp}\phi^{||,\perp}}
\,,
\end{equation}
with
\begin{equation}
\phi_{||} = \frac{3}{2}\int\frac{d\Omega}{4\pi}\,
        \left(\hat{p}_x^2 +\hat{p}_y^2\right)\,\Phi(\vec{p}_f)
\,,
\qquad\qquad
\phi_{\perp} = 3\int\frac{d\Omega}{4\pi}\,
        \hat{p}_z^2\,\Phi(\vec{p}_f)
\,.
\end{equation}
Since $\phi_{||,\perp}\sim\Delta_0^2\sim(1-T/T_c)$ near $T_c$, the
Fermi liquid renormalizations of the penetration depth drop out near
$T_c$, but may give substantial corrections to $\rho_s^{||,\perp}(T)$
at low-temperatures.\cite{leg65,gro86}
\section{Zero-field penetration depth of a {$d_{x^2-y^2}$} superconductor}

Consider the model of the $CuO$ superconductors
based on a $d_{x^2-y^2}$ order parameter. The general form of the order
parameter is,
\begin{equation}\label{dx2-y2}
\Delta(\vec{p}_f)=\Delta (\hat{p}_x^2 - \hat{p}_y^2) * I(\vec{p}_f)
\,,
\end{equation}
where $I(\vec{p}_f)$ is invariant under the full point group, and
$\hat{p}_{x,y}$ define the direction of the Fermi wavevector in the
basal plane of the Fermi surface. Note that the nodes are required by
the broken reflection symmetries. Two parameters determine the
excitation spectrum in the clean limit; (i) the maximum
value of $|\Delta(\vec{p}_f)|$ ($=\Delta_0$), and (ii) the angular
slope of the gap near the node,
\begin{equation}
\mu\equiv\frac{1}{\Delta_0}
\frac{d|\Delta(\vartheta)|}{d\vartheta}\Big|_{\vartheta=\vartheta_{node}}
\,,
\end{equation}
where $\vartheta$ is the angle measured relative to one of the nodes
(see Fig. 2).

\begin{figure}
\centerline{\psfig{figure=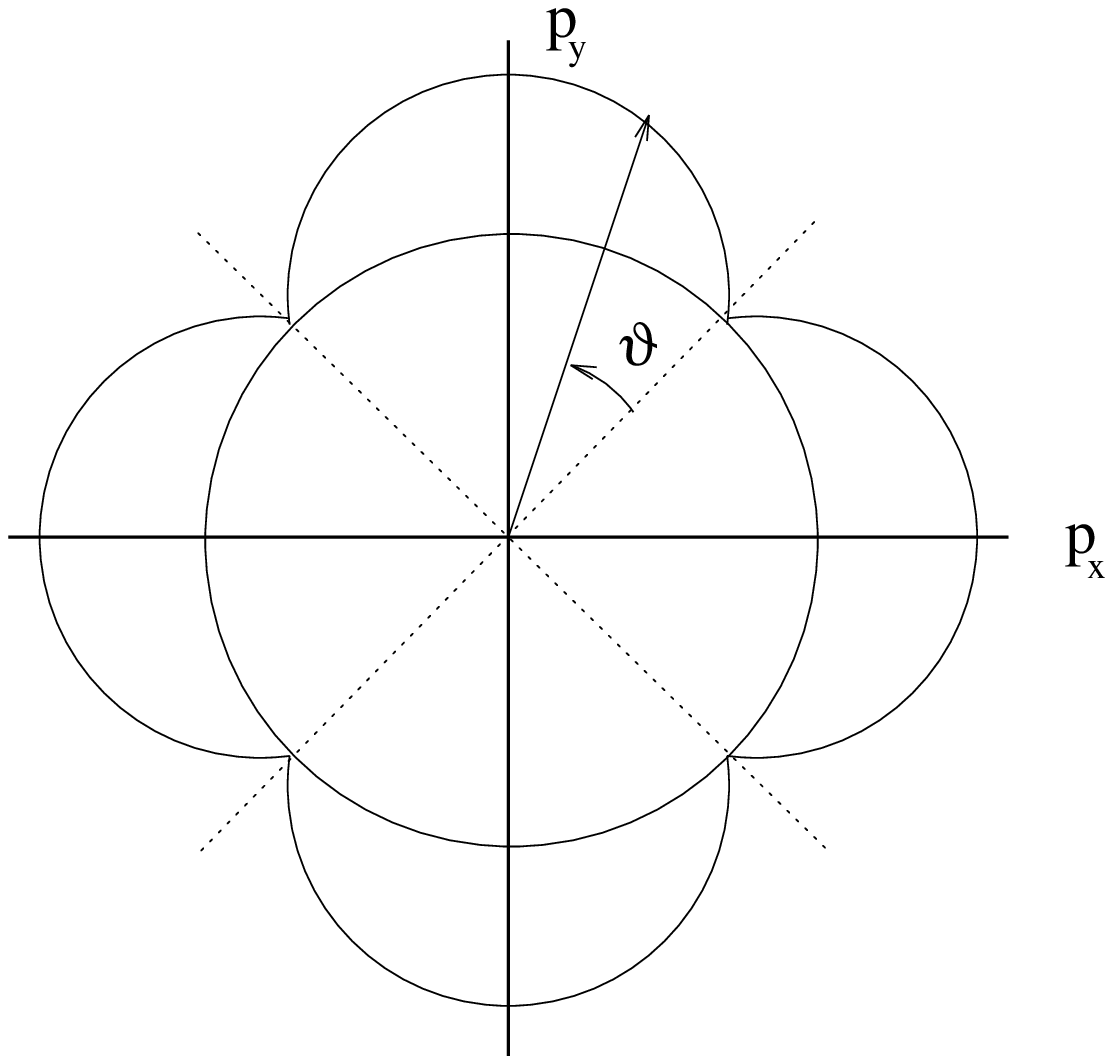,height=2in}}
\begin{quote}
\small
Fig. 2{\hskip 10pt}
Gap function for a $d_{x^2-y^2}$ superconductor.
\end{quote}
\end{figure}

A simple two-parameter model for $|\Delta(\vec{p}_f)|$,
which is useful for numerical calculations, is
\begin{equation}\label{dx2-y2_gap}
|\Delta(\vartheta)|=\Bigg\{
\begin{array}{ll}
& \Delta_0 \qquad ;\qquad \frac{1}{\mu} \le \vartheta \le \frac{\pi}{4} \\
& \mu\Delta_0 \vartheta\qquad ;\quad 0 \le \vartheta < \frac{1}{\mu} \,.
\end{array}
\end{equation}
The maximum gap, for a fixed $\mu$, is obtained from a self-consistent
solution to the gap equation. Figure 3 shows solutions of
eq.(\ref{gap(T)}) for $\Delta_0(T)$ as a function of temperature and
impurity scattering. We obtain a gap ratio of $\Delta_0/T_c = 1.9$ at
$T=0$ for $\mu = 2.7$ and $\Gamma=0$. Note that the leading
temperature-dependent correction to the gap parameter for $T\ll T_c$ is
$\delta\Delta_0(T)\sim T^3$, in the clean limit (appendix A).

\begin{figure}
\centerline{\psfig{figure=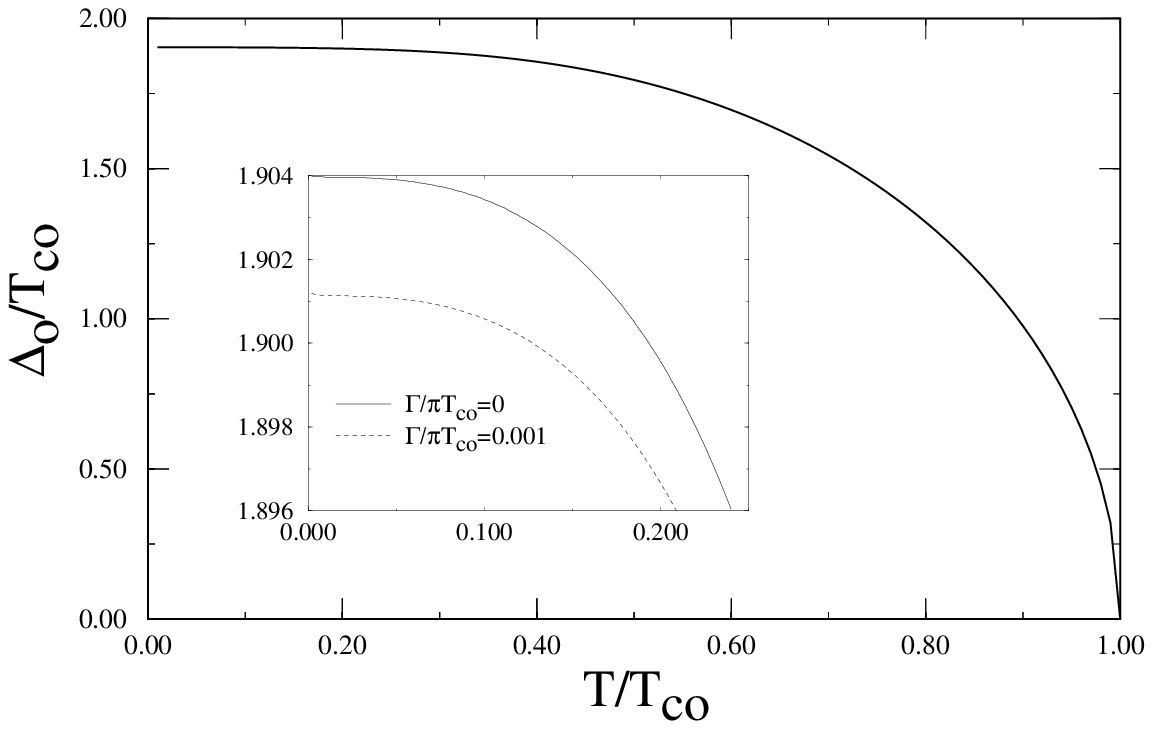,height=2.5in}}
\begin{quote}
\small
Fig. 3{\hskip 10pt}
Maximum gap for a $d_{x^2-y^2}$ superconductor as a function of
temperature. The inset shows the $T^3$ deviation of $\Delta_0(T)$ at
low-temperature.
\end{quote}
\end{figure}

The maximum gap is relatively insensitive to the angular slope of the
gap near the nodes, except for small $\mu$, in which case the nodal
region occupies a significant fraction of phase space. This behavior
can be qualitatively understood by noting that the condensation energy
at zero temperature is effectively determined by the strength of the
pairing interaction, and therefore $T_c$. For an anisotropic gap
$\Delta_0$ is enhanced to compensate for the regions of small gap. If
we assume that the Fermi-surface average of $|\Delta(\vartheta)|^2$ is
constant (fixed by $T_c$), then the maximum gap is given by
$\Delta_0(\mu)\simeq 1.8 T_c/(1 - {8}/{3\pi\mu})$, which
is qualitatively the behavior obtained from the numerical solution to
the gap equation shown in Fig. 4. Note that the ususal one-parameter
$d_{x^2-y^2}$ gap, $\Delta=\Delta_0(\hat{p}_x^2-\hat{p}_y^2)$,
corresponds to $\mu=2$.

\begin{figure}
\centerline{\psfig{figure=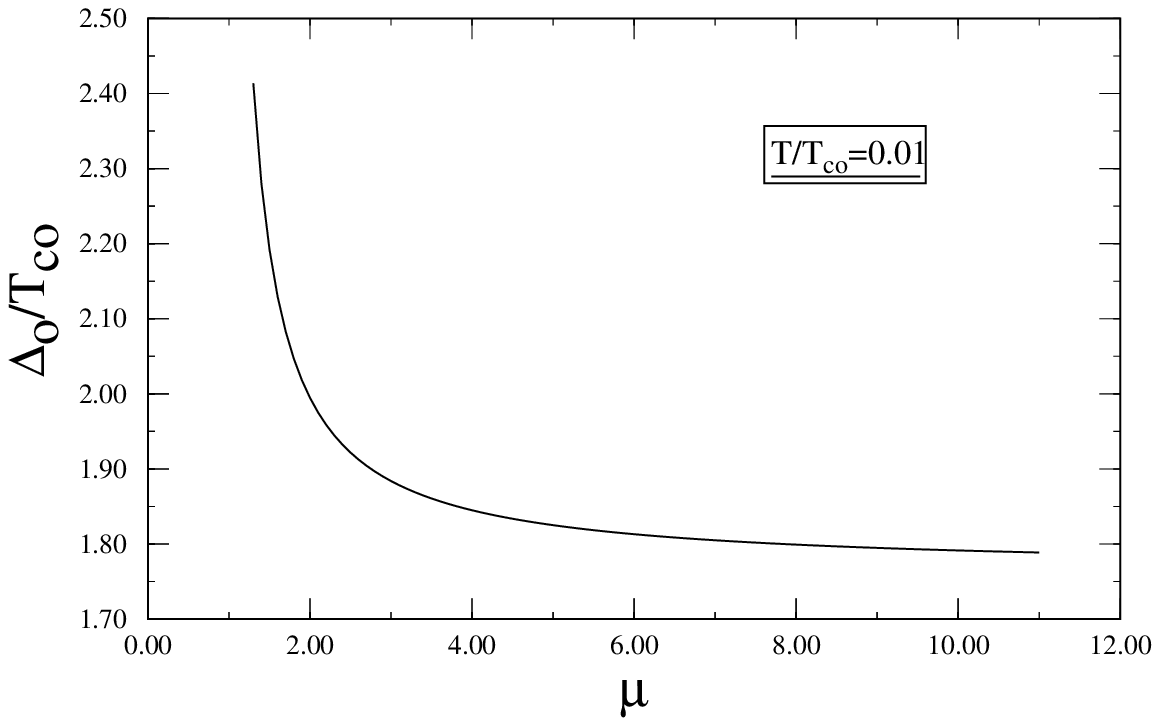,height=2.5in}}
\begin{quote}
\small
Fig. 4{\hskip 10pt}
Maximum gap for a $d_{x^2-y^2}$ superconductor as a function of the angular
slope parameter, $\mu$.
\end{quote}
\end{figure}

The suppression of $T_c$ by impurity scattering is determined by the
pair-breaking parameter $\Gamma=\Gamma_u\sin^2\delta_0$.  Equation
(\ref{Tc_reduction}) for $T_c$ implies that unconventional pairing can
be sensitive to impurity scattering, {\it e.g.\ } the superconducting
transition
is completely suppressed for $\left({\Gamma}/{\pi
T_{co}}\right)_{crit}=\frac{1}{2} e^{-\gamma} \simeq 0.28$, which
corresponds to an impurity mean-free path of $l_{crit}=v_f/2\Gamma
\simeq 3.6 \xi_0$, where $\xi_0=v_f/2\pi T_{c0}$ is the coherence
length in the clean limit.  The relatively small coherence length in
the $CuO$ superconductors is then an advantage for an unconventional
order parameter. For weak pair-breaking, $\Gamma/2\pi T_c \ll 1$, the
suppression of $T_c$ is given by, ${\Delta
T_c}/{T_{c0}}=-{\pi\Gamma}/{8T_{c0}}$. Thus, for $CuO$
superconductors with $T_{c0}=100K$ and a suppression of less than
$0.5K$ we have $\Gamma<1.3K$. For an in-plane coherence length of
$\xi_0=14 \AA$ this corresponds to an impurity mean-free path,
$l>3,450\AA$.\footnote{ For non-magnetic, s-wave impurities in 2D, the
impurity resistivity is given by
$\rho_{imp}^{-1}={e^2}N_fv_f^2/4\Gamma$, where $\Gamma$ is the
same pair-breaking parameter that enters the Abrikosov-Gorkov formula.
Thus, we can express $T_c/T_{c0}$ as a function of $\rho_{imp}$,
independent of the scattering phase shift, $\delta_0$. However,
inelastic scattering, which is important at $T\sim T_c$, destroys this
simple result.\cite{mil88b,rad93}}

The magnitude of the gap parameter is also suppressed by impurity
scattering. Figure 5 shows the suppression of $\Delta_0(0)$ as a
function of $\Gamma$. Note that $\Delta_0(0)$, in contrast to $T_c$, is
more strongly suppressed in the unitarity limit than in the Born
limit for the same suppression of $T_c$.

\begin{figure}
\centerline{\psfig{figure=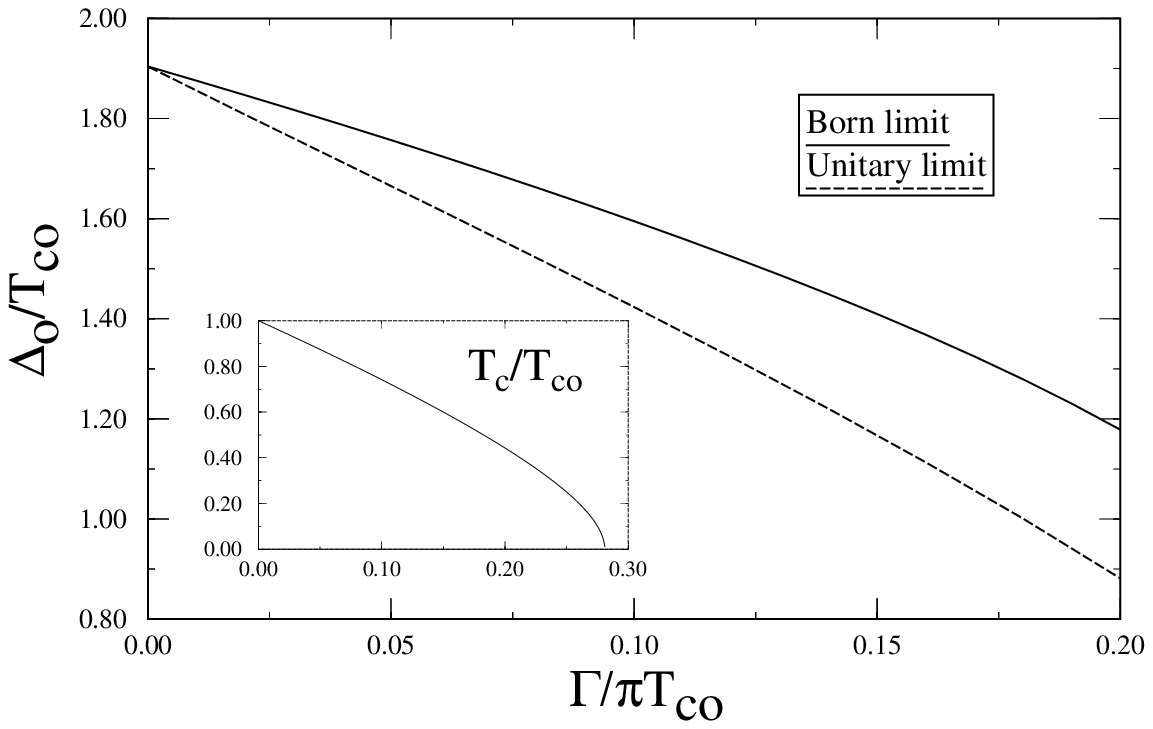,height=2.5in}}
\begin{quote}
\small
Fig. 5{\hskip 10pt}
Maximum gap at $T=0.01 T_c$ for a $d_{x^2-y^2}$ superconductor as a function of
the impurity pair-breaking parameter. The inset shows the impurity
pair-breaking effect on $T_c$.
\end{quote}
\end{figure}

In the clean limit the angle-resolved density of states, obtained from eq.
(\ref{g0}), is given
by the familiar BCS form,
\begin{equation}
{\cal N}(\vec{p}_f,E)=\frac{|E|}{\sqrt{E^2-|\Delta(\vec{p}_f)|^2}}\,
\Theta(E^2-|\Delta(\vec{p}_f)|^2)
\,.
\end{equation}
For low energies, $|E|\ll \Delta_0$, the total density of states is
dominated by the low-lying states near the nodes, and is linear in $|E|$. For
the model gap function
in eq.(\ref{dx2-y2_gap}) the total density of states is
\begin{equation}\label{DOS}
\bar{{\cal N}}(E)=\int\,d\vec{p}_f\,{\cal N}(\vec{p}_f,E)
\simeq\,\frac{2|E|}{\mu\Delta_0}
\,,\qquad|E|<\Delta_0
\,.
\end{equation}
These low-lying excitations are responsible for the linear temperature
dependence of the penetration depth for a superconductor with a line of
nodes (or point nodes in 2D). Not surprisingly, the density of
low-lying states near the nodes is determined by the angular slope of
the gap; therefore, $\mu$ also determines the coefficient of the linear
temperature dependence of the penetration depth (in the clean limit).

Although the small pair size in the cuprates leads to relatively weak
suppression of $T_c$ from impurity scattering, the density of states at
low-energy, $|E|\ll\Delta_0$, and therefore the leading temperature
dependence of the penetration depth, are more sensitive to impurity
scattering, particularly in the strong scattering limit. Figure 6 shows
the density of states as a function of the scattering phase shift for
fixed $\Gamma_u= 0.1\Delta_0$, corresponding to a rather high
impurity concentration. The modification of the density of states at
low-energy is negligible in the Born limit ($\delta_0=\pi/20$) and
remains essentially linear even for intermediate phase shifts
($\delta_0=\pi/4$). However, as the strength of the scattering
increases a finite density of states at $E=0$ develops, becoming of
order $\bar{{\cal N}}(0)\simeq 0.4$ in the unitarity limit. In
addition, $\bar{{\cal N}}(E)$ deviates from linearity below $E\simeq
0.4\Delta_0$. This is the cross-over energy scale, below which impurity
scattering strongly modifies the low-energy spectrum. The cross-over
energy, $\varepsilon^*$, can be calculated from the lowest energy
scale for the renormalized Matsubara frequency $\tilde{\epsilon}_n$ as
$T\rightarrow 0$. From eq. (\ref{Z_imp}) the cross-over scale at $T=0$
is given by
\begin{equation}\label{cross_over_scale}
2\tau=\frac{1}{\sigma\Gamma_u}=\frac{{\cal D}(\varepsilon^*)}
{(1-\sigma)+\sigma(\varepsilon^*\,{\cal D}(\varepsilon^*))^2}
\,,
\end{equation}
where $\sigma=\sin^2\delta_0$, and ${\cal D}(\varepsilon^*)$ can
be calculated to leading logarithmic accuracy ($\ln(\Delta_0/\varepsilon^*)\gg
1$) for the $d_{x^2-y^2}$ state,
\begin{equation}
{\cal D}(\varepsilon^*)\simeq
\frac{4}{\pi\mu\Delta_0}\,\ln(2\Delta_0/\varepsilon^*)
\,.
\end{equation}
In the unitarity
limit this scale can be a sizable fraction of $\Delta_0$ even in the
dilute limit, $\varepsilon^*\sim \sqrt{\Gamma_u\Delta_0}$, but in the
Born limit the cross-over scale is exponentially small,
$\varepsilon^*\sim 2\Delta_0\,\exp({-\mu\pi\tau\Delta_0/2})$. For more detailed
discussions of the density of states of d-wave superconductors see Refs.
(\cite{pre94,feh94}).

\begin{figure}
\centerline{\psfig{figure=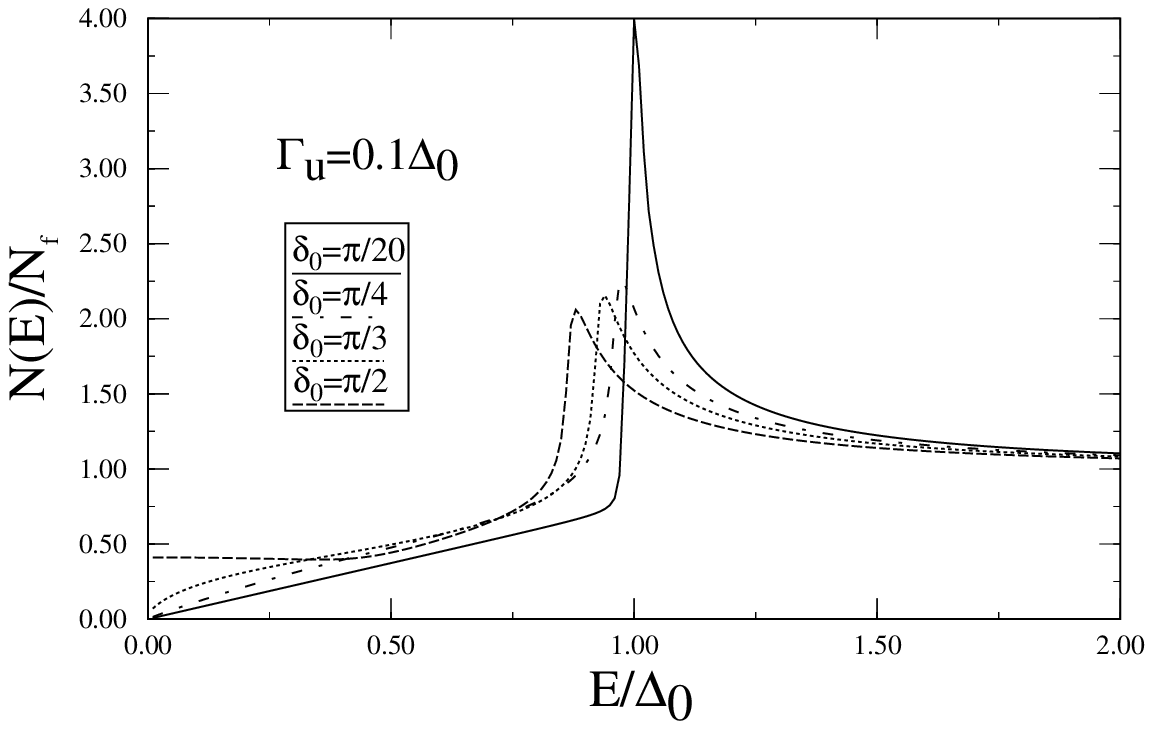,height=2.5in}}
\begin{quote}
\small
Fig. 6{\hskip 10pt}
Density of states vs. scattering phase shift. The impurity
concentration is fixed with $\Gamma_u/\Delta_0=0.1$. Note the finite
density of states and the suppression of the maximum gap in the
unitarity limit.
\end{quote}
\end{figure}

\subsection*{Penetration depth for the $d_{x^2-y^2}$ gap}

The temperature dependence of the penetration depth for a $d_{x^2-y^2}$
gap function is obtained from eq.(\ref{rhos}) (neglecting Fermi liquid
corrections). We assume a cylindrical Fermi surface and a gap function
parametrized by $\mu$ and $\Delta_0$ as in eq. (\ref{dx2-y2_gap}). The
in-plane penetration depth becomes
\begin{equation}\label{lambda}
\frac{1}{\lambda_{\|}^2}=\frac{4\pi e^2 N_f (v_f^{\|})^2}{c^2}\,
\oint\frac{d\vartheta}{2\pi}\,
\pi T\sum_{\epsilon_n}\frac{|\Delta(\vartheta)|^2}
{[Z(\epsilon_n)^2\epsilon_n^2+|\Delta(\vartheta)|^2]^{3/2}}
\,,
\end{equation}
where the integration is over the Fermi circle in the basal plane, and
$Z(\epsilon_n)$ is the impurity renormalization factor for s-wave
scattering centers. In the clean limit the leading correction to the
penetration depth at low-temperatures is linear in $T$, typical of a gap with a
line of nodes,\cite{gro86,cho88,pro91,hir93}
\begin{equation}\label{linearlambda}
\frac{\delta\lambda_{\|}(T)}{\lambda_{\|}(0)} =
\left(\frac{2\ln 2}{d|\Delta(\vartheta)|/d\vartheta|_{node}}
\right) \, T + ...\qquad ; \quad T \ll T_c
\,,
\end{equation}
where $d|\Delta(\vartheta)|/d\vartheta|_{node}$ is the angular slope of
$|\Delta(\vartheta)|$ at the node. For the model gap in eq.
(\ref{dx2-y2_gap}),
$d|\Delta(\vartheta)|/d\vartheta|_{node}=\mu\Delta_0$.

The report by Hardy, {{\it et al.\ }}\cite{har93} of a linear temperature
dependence to $\delta\lambda_{||}(T)$ for single crystals of
YBa$_{2}$Cu$_{3}$O$_{6.95}$ provided the first substantial evidence for
a superconducting state with a line of nodes in the excitation gap.
Figure 7 shows the low-temperature data for $\delta\lambda_{||}(T)$
reported in Ref.(\onlinecite{son94}). The solid line is a calculation
of the penetration depth for a $d_{x^2-y^2}$ gap with the angular slope
adjusted to fit the data for $T<20\,K$. For the absolute penetration
depth we assume $\lambda_{||}(0)=1,400\,\AA$.\cite{har89} The slope of
the penetration depth at low temperature is $4.3\,\AA/K$, and the fit
to eq.(\ref{linearlambda}) gives $\mu=2.7$, $\Delta_0(0)/T_c=1.9$, and
$d|\Delta(\vartheta)|/d\vartheta|_{node}=\mu\Delta_0(0)\simeq
5.1\,T_c$. For comparison, the one-parameter $d_{x^2-y^2}$ model,
$\Delta=\Delta_0\cos(2\phi)$ with $\Delta_0$ calculated
self-consistently, gives
$d|\Delta(\vartheta)|/d\vartheta|_{node}=\mu\Delta_0(0)=4.3\,T_c$.

\begin{figure}
\centerline{\psfig{figure=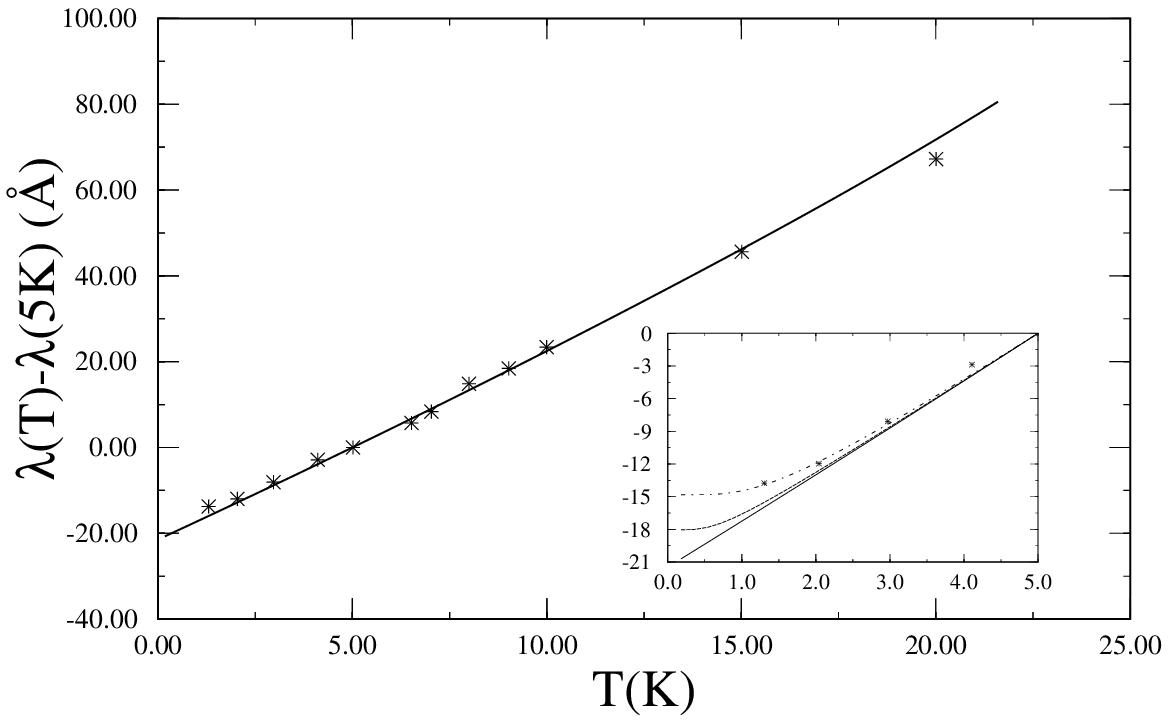,height=2.5in}}
\begin{quote}
\small
Fig. 7{\hskip 10pt} Penetration depth as a function of temperature. The
(*) are experimental data obtained from Ref. (\onlinecite{har93}) the
solid line is the best fit with $\mu\simeq 2.7$, $\Delta_0/T_c\simeq
1.9$. The inset shows the effect of a tiny gap at the nodal positions
on the temperature dependence of the penetration depth. The axes are
the same as those of the main graph. The middle (top) curve
in the inset corresponds to a tiny gap of $\Delta_{min}/\Delta_{0}=1\%$
($2\%$).
\end{quote}
\end{figure}

Also shown in Fig. 7 is the effect of a tiny gap at the nodal
positions. Tiny gaps can arise in strongly anisotropic conventional
superconductors,\cite{cha93} or if the unconventional order parameter
contains a small component of another representation. For example, an
order parameter of the form $d_{x^2-y^2} + i \sqrt{\varepsilon} d_{xy}$
has a gap function, $|\Delta(\vec{p}_f)|= \Delta_0[\cos^2(2\phi) +
\varepsilon\sin^2(2\phi)]^{1/2}$, which is strongly anisotropic for
small $\varepsilon$ with a tiny gap of order $\varepsilon\Delta_0$ near
the nodes of the $d_{x^2-y^2}$ component.\cite{lau94} The data of Ref.
(\onlinecite{har93}) implies that $\varepsilon<2\,\%$.

Precision measurements of the penetration depth in thin films show a
$T^2$ behavior at low temperatures rather than the linear temperature
dependence characteristic of a $d_{x^2-y^2}$ order
parameter.\cite{fio88,ann91,pon91,anl92,ach93,lee93a,lee94} This
difference may be due to scattering by a higher concentration of
defects present in the films. Hirschfeld and Goldenfeld\cite{hir93}
argue that the $T^2$ dependence in films and the $T$ dependence of
$\delta\lambda_{||}(T)$ reported for single crystals can be understood
within the same $d_{x^2-y^2}$ model for the pairing state provided the
films are relatively dirty compared to the single crystals. However,
impurity scattering or defect scattering is pair-breaking in
unconventional superconductors, so in order to explain the weak or
negligible suppression of $T_c$ in films (compared to $T_c$ in the
single crystals of Ref.(\onlinecite{har93}), the authors of Ref.
(\onlinecite{hir93}) argue that the scattering responsible for the
$T^2$ dependence of $\delta\lambda_{||}(T)$ results from a {\it dilute}
concentration of strong scattering centers with phase shifts near the
unitarity limit. A small concentration of unitarity scatterers leads to
strong modification of the density of states in the small phase space
region near the nodes. Since $\delta\lambda_{||}$ is determined by
these low-energy excitations, the temperature dependence of
$\delta\lambda_{||}(T)$ for $T\ll T_c$ may be strongly modified even
when the suppression of $T_c$ by a dilute concentration of scatterers
is negligible.

In the Born limit (weak scattering) the density of states, even near
the nodes is nearly unchanged. Thus, a much higher concentration of
defects is needed to generate $\delta\lambda_{||}(T)\sim T^2$ for
$T\lesssim 0.2 T_c$, which is accompanied by a sizeable suppression of
$T_c$. The sensitivity of the low-energy excitation spectrum to the
scattering strength is reflected in the temperature dependence of the
penetration depth shown in Fig. 8. The concentration of s-wave
scatterers is fixed and the curves show the evolution from a linear $T$
dependence in the Born limit ($\delta_0=\pi/20$), and intermediate
phase shifts, to the $T^2$ dependence in the unitarity limit
($\delta_0=\pi/2$). Note that the cross-over from
$\delta\lambda_{||}\sim T$ to $\delta\lambda_{||}\sim T^2$ is abrupt,
occuring very near the unitarity limit for dilute concentrations. A
sharp cross-over in the excitation spectrum as a function of phase shift
has also been noted by Preosti, {\it et al.\ } \cite{pre94}. Thus, it is
worth emphasizing that for dilute point impurities it is not merely
`strong scattering' that is required to obtain $\delta\lambda_{||}\sim
T^2$ with minimal reduction in $T_c$, but scattering with
$\delta_0\rightarrow\pi/2$.

\begin{figure}
\centerline{\psfig{figure=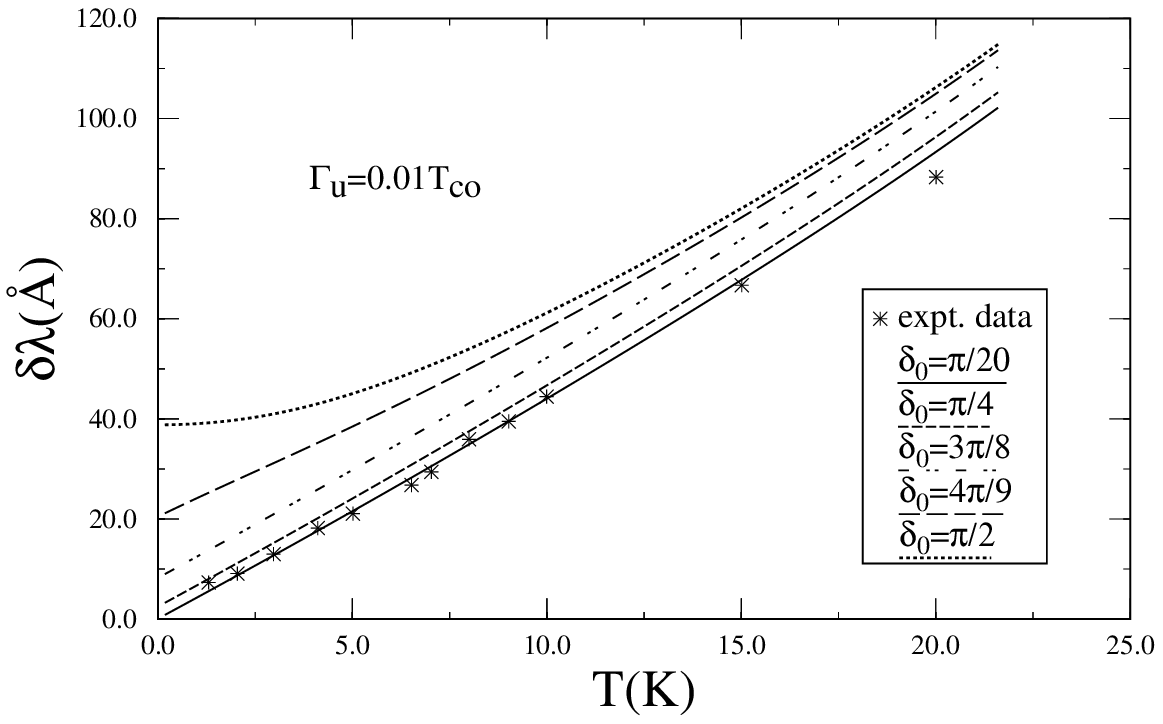,height=2.5in}}
\begin{quote}
\small
Fig. 8{\hskip 10pt}
Penetration depth as a function of temperature for phase shifts ranging
from the Born limit to the unitarity limit. The concentration is fixed
with $\Gamma_u=0.01\,T_c$.
\end{quote}
\end{figure}

In the unitarity limit even dilute concentrations of impurities can
strongly modify the low-energy spectrum. In the unitarity limit the
cross-over energy scale is $\varepsilon^*\sim \sqrt{\Gamma_u\Delta_0}$.
Thus, $\delta\lambda_{||}(T)$ deviates from the linear $T$ dependence
for $T<\varepsilon^*$. The sensitivity of $\delta\lambda_{||}(T)$ at
$T\ll T_c$ for unitarity scattering (see Fig. 9) places a strong
constraint on the concentration of scatterers that can be present in
clean single crystals that shows $\delta\lambda_{||}(T)\sim T$ down to
low temperatures. For the data of Ref. (\onlinecite{har93}) we find
$\Gamma_u/\Delta_0\lesssim 0.0001$ given that
$\delta\lambda_{||}(T)\sim T$ down to $T\simeq 1\,K$.

\begin{figure}
\centerline{\psfig{figure=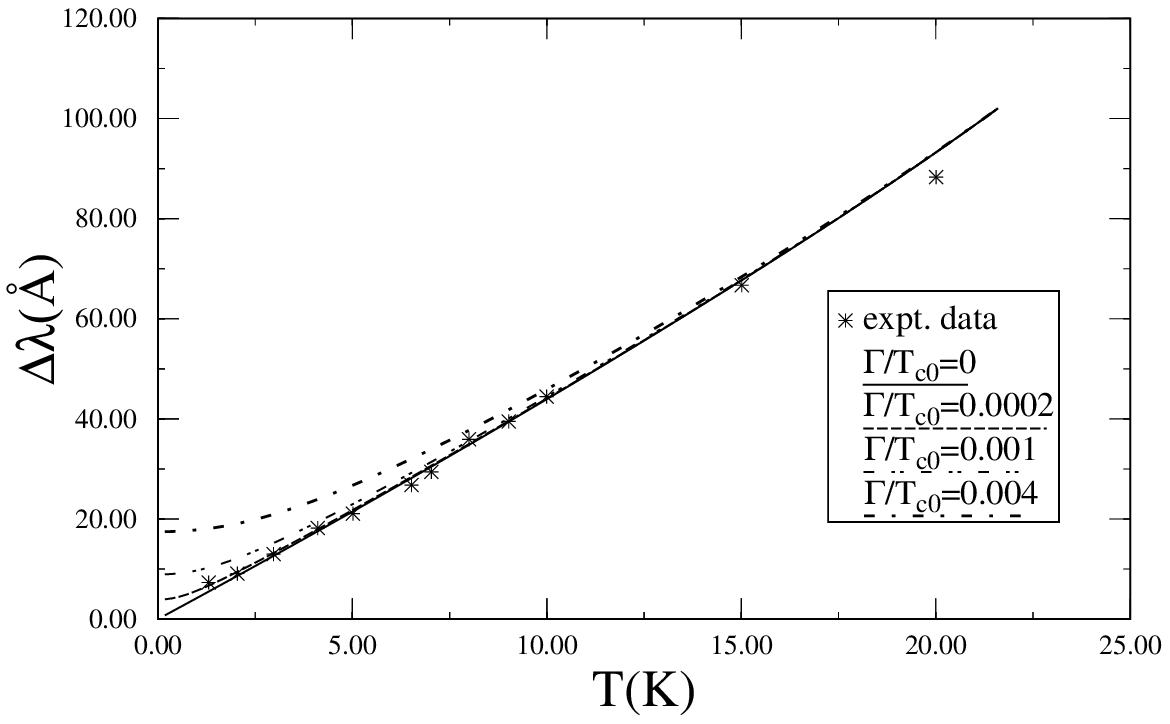,height=2.5in}}
\begin{quote}
\small
Fig. 9{\hskip 10pt}
Penetration depth as a function of temperature and impurity concentration
($\sim\Gamma_u$) in the unitarity
scattering limit.
\end{quote}
\end{figure}

\section{Nonlinear Current Response}

In the Meissner geometry the screening current is proportional to the
applied surface field, $j_s\sim cH/\lambda$. As $H$ is increased
nonlinear field corrections to the constitutive equation for the
supercurrent may become significant. In conventional type II
superconductors nonlinear corrections to the current-velocity relation
arise from the thermal population of quasiparticles, and vortex
nucleation generally occurs before these nonlinear effects become
important.

In unconventional superconductors with nodes in the excitation gap the
nonlinear field correction to the supercurrent is substantially larger
than in conventional superconductors with other similar material
properties. The origin of the anomalous nonlinear Meissner effect is
the contribution to the screening current associated with the
quasiparticle states near the nodal lines. As a result the nonlinear
Meissner effect may be used to detect the nodal structure of the gap of
an unconventional superconductor.\cite{yip92a}

In the limit $|{\vec{v}_f\cdot\vec{\nabla}|\Delta|}|/{\pi T_c}\sim
\xi/\lambda\ll 1$ the  current can be expressed as a local function of
the condensate velocity, $\vec{v}_s(\vec{R})$. In the presence of a
condensate flow the local solution to the transport equation,
$\hat{g}(\vec{p}_f,\vec{R};\epsilon_n)$, is given by eq. (\ref{g0}),
but evaluated with $i\tilde{\epsilon}_n \rightarrow i\tilde{\epsilon}_n
- \sigma_v$, where $\sigma_v(\vec{p}_f,\vec{R}) =
\vec{v}_f\cdot\vec{v}_s(\vec{R})$. The current response obtained from
eq. (\ref{QCcurrent}) is
\begin{equation}\label{current_mats}
\vec j_s =
-2eN_f\int d\vec{p}_f\,\vec{v}_f(\vec{p}_f)\,\pi T\sum_{\epsilon_n}\,
{{\sigma_v(\vec{p}_f)- i\tilde{\epsilon}_n}
\over
{\sqrt{(\tilde{\epsilon}_n+i\sigma_v(\vec{p}_f))^2
 +|\tilde{\Delta}(\vec{p}_f;\epsilon_n)|^2}}}
\,.
\end{equation}
One point to note in calculating the density of states, or current, at
finite flow with impurity scattering is that the
impurity-renormalization of $\Delta(\vec{p}_f)$ need not vanish, even
for s-wave impurity scattering and $\left<\Delta(\vec{p}_f)\right>=0$.
The reason is clear from eq. (\ref{renormdelta}) for
$\tilde{\Delta}(\vec{p}_f;\epsilon_n)$; the kernel no longer vanishes
by symmetry with the replacement, $\tilde{\epsilon}_n\rightarrow
\tilde{\epsilon}_n+i\vec{v}_f\cdot\vec{v}_s$ for a general flow field.
However, for special directions of $\vec{v}_s$, {\it e.g.\ } $\vec{v}_s$
parallel to a node, the impurity correction to $\Delta(\vec{p}_f)$
vanishes, $\tilde{\Delta}=\Delta(\vec{p}_f)$.

Equation (\ref{current_mats}) for the current can be transformed by
contour integration and analytic continuation to the
real axis to give
\begin{equation}\label{current_realaxis}
\vec{j}_s = -2eN_f
\int_{\sigma_v>0}\,d\vec{p}_f\,\vec{v}_f(\vec{p}_f)\,
\int_{-\infty}^{+\infty}\,dE\,f(E)\,
\left[{\cal N}_{+}(\vec{p}_f,E) - {\cal N}_{-}(\vec{p}_f,E)\right]
\,,
\end{equation}
where ${\cal N}_{\pm}(\vec{p}_f,E)$ is the density of states for
quasiparticles that are co-moving ($+\vec{v}_f\cdot\vec{v}_s>0$) and
counter-moving ($-\vec{v}_f\cdot\vec{v}_s<0$) relative to the condensate
flow. The integral is taken over the half space
$\sigma_v=\vec{v}_f\cdot\vec{v}_s>0$ with the counter-moving
excitations included by inversion symmetry: ${\cal
C}_i\sigma_v=-\sigma_v$ and $|\Delta({\cal
C}_i\vec{p}_f)|=|\Delta(\vec{p}_f)|$. This result is general enough to
cover nonlinear field corrections to the current for superconductors
with an unconventional order parameter and pair-breaking effects from
impurity scattering.

\begin{figure}
\centerline{\psfig{figure=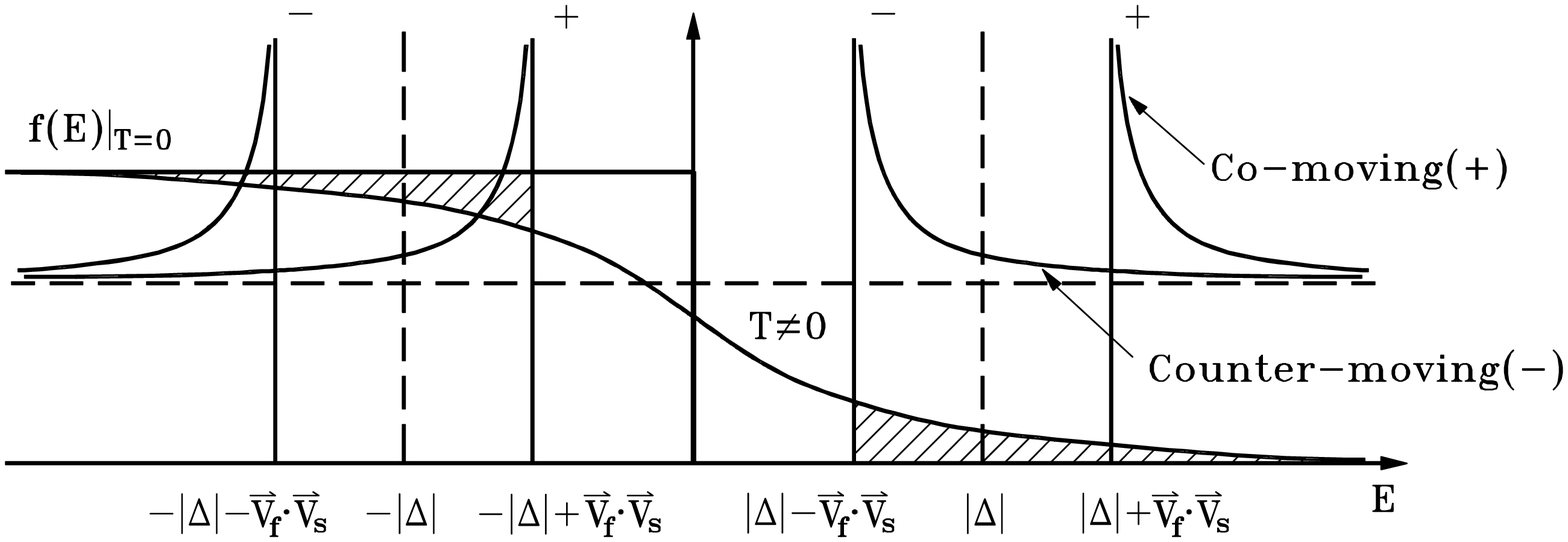,width=5in}}
\begin{quote}
\small
Fig. 10{\hskip 10pt}
Density of states for co-moving ($+\vec{v}_f\cdot\vec{v}_s$) and
counter-moving ($-\vec{v}_f\cdot\vec{v}_s$) excitations at a point
$\vec{p}_f$ on the Fermi-surface where
$|\vec{v}_f\cdot\vec{v}_s|<|\Delta(\vec{p}_f)|$.
\end{quote}
\end{figure}

The difference in the nonlinear current-velocity relation for
conventional and unconventional order parameters appears in the
contributions to the current from the co-moving and counter-moving
excitation spectrum at $T=0$. The spectrum is shown Fig. 10 in the clean
limit for a specific direction $\vec{p}_f$ in which
$\vec{v}_f\cdot\vec{v}_s<|\Delta(\vec{p}_f)|$. At zero temperature only
the co-moving and counter-moving quasiparticle states with $E<0$
contribute to the current.

\subsection*{Nonlinear current: conventional gap}

For a conventional superconductor with an isotropic gap at $T=0$ the
current is easily calculated from the difference in the number of
co-moving versus counter-moving quasiparticles that make up the
condensate,
\begin{equation}
\vec{j}_s
=
-2eN_f \int_{\sigma_v>0}\,d\vec{p}_f\,\vec{v}_f(\vec{p}_f)\,
\left[2\vec{v}_f\cdot\vec{v}_s\right]
=
-e \rho\,\vec{v}_s
\quad , \quad v_s < v_f/\Delta_0
\,,
\end{equation}
with $\rho=N_f v_f^2$ for a cylindrical Fermi surface. The
main point is that the current is linear in $\vec{v}_s$ for velocities
up to the bulk critical velocity, $v_c=v_f/\Delta_0$. At $v_s = v_c$
the edge of the spectrum for the upper branch ($E>0$ for $v_s < v_c$)
of counter-moving excitations drops below $E=0$, and the edge of the
spectrum for the lower branch ($E<0$ for $v_s < v_c$) of co-moving
excitations shifts above $E=0$. As a result the current carried by the
condensate drops rapidly above the critical velocity. The current is
nonanalytic at $v_c$ because a branch of counter-moving (co-moving)
excitations that are unoccupied (occupied) for $v_s<v_c$ become
occupied (unoccupied) $v_s>v_c$. For example, for a 1D Fermi surface
the current becomes, $j_s = -2eN_f v_f^2
\left\{v_s - \Theta(v_s-v_c)\sqrt{v^2_s - v^2_c}\right\}$.

At non-zero temperatures thermal occupation of the upper branches and
de-population of the lower branches reduces the condensate
supercurrent. In the clean limit eq.(\ref{current_realaxis}) can be
transformed to
\begin{equation}\label{js_clean}
\vec{j}_s = -2eN_f\,\int d\vec{p}_f\,\vec{v}_f\,(\vec{v}_f\cdot\vec{v}_s)
+ \vec{j}_{qp}
\,,
\end{equation}
\begin{equation}\label{qpcurrent}
\vec{j}_{qp} = - 4eN_f
\int_{0}^{\infty}\,d\xi
\int\,d\vec{p}_f\,\vec{v}_f\,
\left[f\left(\sqrt{\xi^2+|\Delta|^2}
+\vec{v}_f\cdot\vec{v}_s\right)\right]
\,,
\end{equation}
which separates the condensate contribution to the current, the fully
occupied negative energy branches shown in Fig. 10, from $\vec{j}_{qp}$,
the current carried by the excitations associated with population of
the upper branches and de-population of the lower branches in Fig. 10.
The current carried by the excitations is a backflow current. For low
velocities the net current is linear in $\vec{v}_s$,
$\vec{j}_s=-e\rho_s(T)\,\vec{v}_s$, where $\rho_s(T)$ is the superfluid
density in the two-fluid model,
\begin{equation}
\rho_s(T)=\rho - N_f v_f^2 \int_{0}^{\infty}\,d\xi
\frac{{\rm sech}^2(\sqrt{\xi^2+|\Delta|^2})}{2T}
\,.
\end{equation}
For larger flow velocities the linear relation breaks down. The leading
correction to the current-velocity relation for $v_s<v_c$ is,
\begin{equation}
\vec{j}_s=-e\rho_s(T)\ \vec{v}_s\,
\left[1-\alpha(T)\left ({v_s\over v_c}\right )^2\right]
\label{current_swave}
\,.
\end{equation}
The nonlinear correction is third-order in $v_s$ and determined by the
coefficient $\alpha(T)\ge 0$ and the bulk critical velocity,
$v_c=\Delta(T)/v_f$.  There are two sources to the nonlinear current.
At finite $v_s$ there is a difference in the thermal occupations of
co-moving and counter-moving excitations. Since the counter-moving
branch of has a larger occupation than the co-moving branch ($f(E-v_fv_s)$
compared to
$f(E+v_fv_s)$), the thermal excitations
further reduce the current compared with the linear response value. In
addition the condensate velocity is also pairbreaking, {\it i.e.\ }
$v_s$ reduces the magnitude of the mean-field gap parameter, further
reducing the current density at finite flow. Figure 11 shows the
temperature dependence of the nonlinear correction to
eq.(\ref{current_swave}).  Note that $\alpha(T)\sim\exp(-\Delta/T)$ for
$T\rightarrow 0$. There are no excitations that contribute to the
backflow current at $T=0$ and the constitutive equation is strictly
linear for velocities below the bulk critical velocity,
$v_c=\Delta/v_f$.

\begin{figure}
\centerline{\psfig{figure=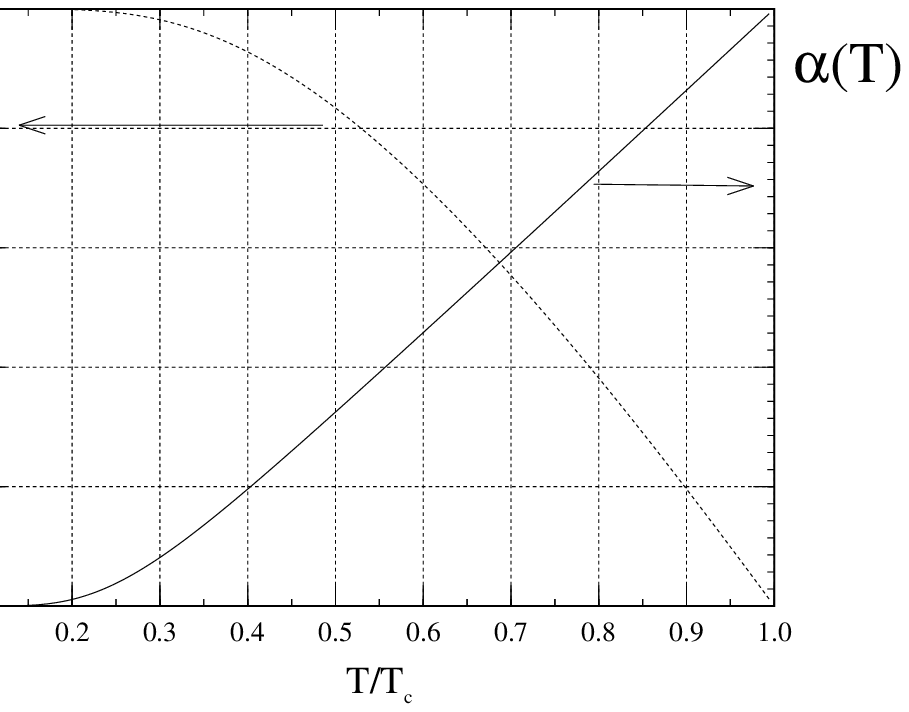,height=2.5in}}
\begin{quote}
\small
Fig. 11{\hskip 10pt}
Temperature dependence of $\rho_s(T)$ and the nonlinear coefficient $\alpha(T)$
for an s-wave gap.
\end{quote}
\end{figure}

The relevance of the nonlinear current-velocity constitutive equation
to the penetration of magnetic fields into a superconductor is
qualitatively clear. The reduction of the current by the flow reduces
the effective superfluid density, and, therefore, increases the
penetration of the field into the superconductor. Since the current is
proportional to the field in linear order, the leading correction to
the effective penetration length is quadratic in the surface field.
Solution of the nonlinear London equation in the Meissner geometry
gives the following result for the field-dependence of the penetration
depth (see appendix B),
\begin{equation}
{1\over\lambda(T,H)}={1\over\lambda(T)}\
\left\{
1-{3\over 4}\alpha(T)\left [{H\over H_{0}(T)}\right ]^2
\right\}
\end{equation}
where $1/\lambda(T)^2={8\pi}e^2\rho_s(T)/3c^2$ is the zero-field London
penetration depth, and $H_{0}(T)=e\ \lambda(T)/c\ v_c(T)$ is of order
the thermodynamic critical field. Thus, in the London limit nonlinear
Meissner effect in a conventional superconductor is exponentially small
at low temperatures and is quadratic in $H/H_0$.

\subsection*{Nonanalytic supercurrents at T=0: $d_{x^2-y^2}$ gap}

The expansion of the current in $\vec{v}_s$ breaks down for an
unconventional superconductor with nodes in the gap. This is clear from
eq. (\ref{current_mats}); a Taylor expansion in $\sigma_v$ fails for
$T\ll T_c$ when there are directions $\vec{p}_f$ where
$|\Delta(\vec{p}_f)|$ and $|\tilde{\epsilon}_n|$ are always small
compared to $|\sigma_v(\vec{p}_f)|$. In the clean limit for a
$d_{x^2-y^2}$ gap the breakdown of the Taylor expansion leads to a
nonanalytic current-velocity relation at $T=0$ of the form
$\vec{j}_s=-e\rho\,\vec{v}_s\,\{1-{|\vec{v}_s|}/{v_0}\}$, where
$v_0\sim\Delta_0/v_f$.\footnote{Nonanalytic currents have been
investigated in superfluid $^3$He-A, initially by Volovik and
Mineev.\cite{vol81b} This work is closely related to a number of
theoretical investigations of the hydrodynamical equations of
superfluid $^3$He in the limit $T\rightarrow 0$ (see Ref.
(\cite{com86}). For an analysis of the non-analytic current in $^3$He-A
see Ref. (\cite{muz83}); these authors also calculate the non-analytic
current-velocity relation for an axially symmetric, polar state with
$\Delta\sim \hat{p}_z$. The polar model is examined in more detail by
Choi and Muzikar.\cite{cho87}} The physical origin of this nonanalytic
current is easily understood by considering the $d_{x^2-y^2}$ gap with
$\vec{v}_s$ directed along a node, as shown in Fig. 12 (left panel).

\begin{figure}
\centerline{\psfig{figure=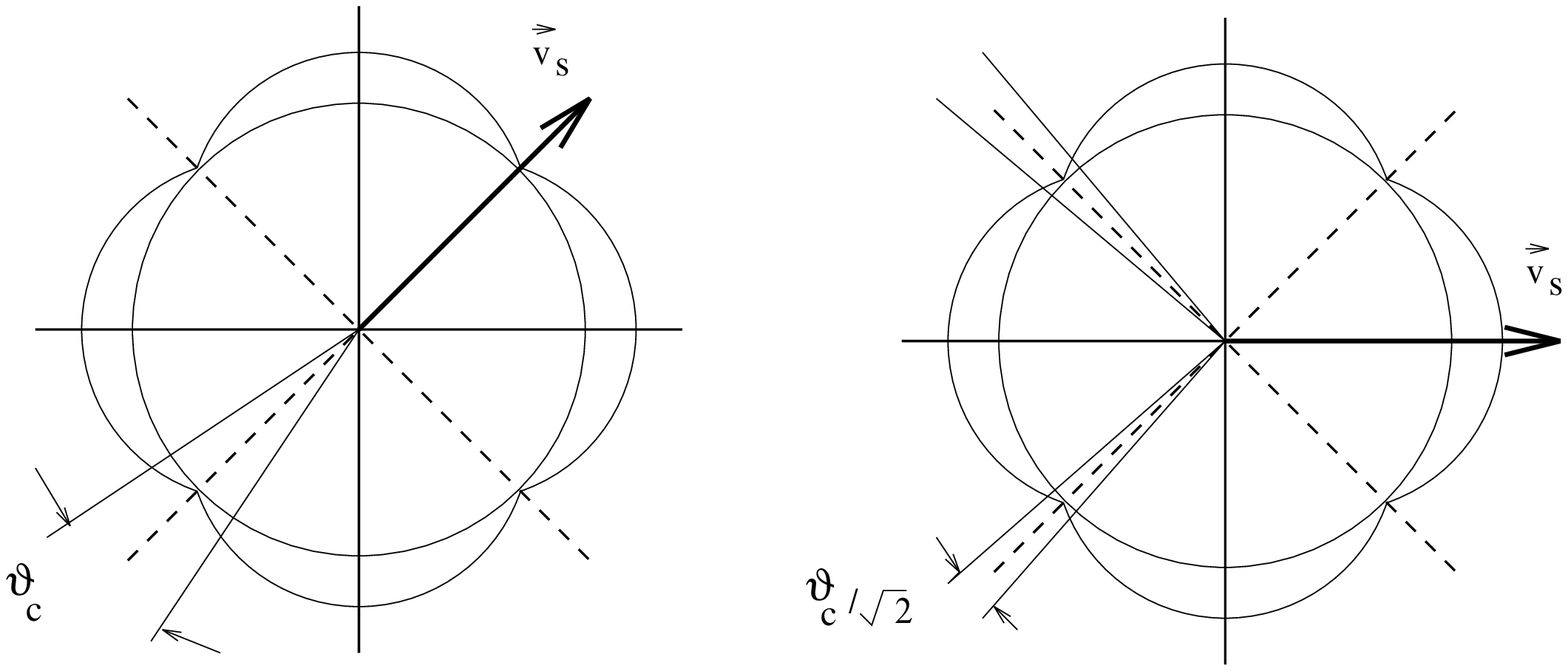,height=2.0in}}
\begin{quote}
\small
Fig. 12{\hskip 10pt}
Phase space contributing to the quasiparticle backflow jets at $T=0$
for $\vec{v}_s ||$ node, and $\vec{v}_s ||$ antinode.
\end{quote}
\end{figure}
For any $v_s\ne 0$ there is a region of the Fermi surface with
$|\Delta(\vec{p}_f)|+\vec{v}_f\cdot\vec{v}_s < 0$, in which the upper
branch of the counter-moving excitations (see Fig. 10) have negative
energy and become populated. The non-analytic dependence on $\vec{v}_s$
reflects the occupation of this counter-moving branch of excitations at
$T=0$. Figure 12 illustrates the phase space contributing to the
backflow current. For $|v_f v_s|\ll\Delta_0$ the wedge of occupied
states is $-\vartheta_c\le\vartheta\le\vartheta_c$, with
$\vartheta_c=v^*_f v_s/\mu\Delta_0$, where $v_f^*$ is the Fermi
velocity at the node and $\mu\Delta_0$ is the angular slope of
$|\Delta(\vartheta)|$ at the node. The current is calculated by
transforming eq. (\ref{current_realaxis}) to
\begin{equation}
\vec{j}_s = -2eN_f\,\int d\vec{p}_f\,\vec{v}_f
\left[(\vec{v}_f\cdot\vec{v}_s)
+
2\int_{0}^{\infty}\,d\xi\,
f(E(\xi)+\vec{v}_f\cdot\vec{v}_s)\right]
\,,
\end{equation}
where $E(\xi)=\sqrt{\xi^2+|\Delta(\vec{p}_f)|^2}$. The first term is
the condensate current, $-e\rho\vec{v}_s$. The backflow current at
$T=0$ is easily calculated from the phase space of occupied
counter-moving excitations. With the velocity directed along the nodal
line $\hat{p}_x=\hat{p}_y$, {\it i.e.\ } $\vec{v}_s=v_s\hat{x}'$ as shown in
Fig. 12 (left panel), the occupied states give
\begin{eqnarray}\displaystyle
\vec{j}_{qp}=-2eN_f\int d\vec{p}_f \vec{v}_f\,
\Theta(-\vec{v}_f\cdot\vec{v}_s-|\Delta(\vec{p}_f)|)\,
\sqrt{(\vec{v}_f\cdot\vec{v}_s)^2 - |\Delta(\vec{p}_f)|^2} \\
\nonumber
=-2eN_f\,v^{*}_f\,\int_{-\vartheta_c}^{\vartheta_c}\,
\frac{d\vartheta}{2\pi}\,\sqrt{(v_f^*v_s)^2-(\mu\Delta_0\vartheta)^2}
\,(-\hat{x}')
\,.
\end{eqnarray}
To leading order in $({v^*_f v_s/\Delta_0})$ we obtain a
total current of
\begin{equation}
\vec{j}_s=-e\rho\vec{v}_s\,
\left\{1-\frac{|\vec{v}_s|}{v_0}\right\}
\qquad(\vec{v}_s||\,{\rm node})\,,
\label{cur_node}
\end{equation}
where $v_0=\mu\Delta_0/v^*_f$ is of order the bulk critcial velocity
scale.  Equation (\ref{cur_node}) clearly holds for $\vec{v}_s$
directed along any of the four nodes.  Note that the current is
parallel to the velocity, and that the counter-moving excitations
reduce the supercurrent, as expected. Also the nonlinear correction is
quadratic rather than cubic, as is obtained for the conventional gap,
and with the characteristic scale determined by
$v_0=\mu\Delta_0/v^*_f$.

Unlike the linear response current, the nonlinear quasiparticle current
is anisotropic in the basal plane. A velocity field directed along the
maximum direction of the gap (antinode), $\vec{v}_s=v_s\ \hat{x}$,
produces two counter-moving jets, albeit with reduced magnitude because
the projection of $\vec{v}_s$ along the nodal lines is reduced by
$1/\sqrt{2}$ (Fig. 12, right panel).
The critical angle defining the occupied states in this
case is given by $\frac{1}{\sqrt{2}}(v_s/v_0)$.
The resulting current is
easily calculated to be
\begin{equation}
\vec{j}_{qp}=-e\rho\left[
\frac{v_s }{\sqrt{2}}\left(\frac{v_s/\sqrt{2}}{v_0}\right)(-\hat{x}')
+
\frac{v_s }{\sqrt{2}}\left(\frac{v_s/\sqrt{2}}{v_0}\right)(+\hat{y}')
\right]
\,,
\end{equation}
giving a total current of
\begin{equation}
\vec{j}_s=-e\rho\,\vec{v}_s\,
\left\{1-\frac{1}{\sqrt{2}}\,\frac{|\vec{v}_s|}{v_0}\right\}
\qquad(\vec{v}_s||\,{\rm antinode})
\,,
\label{cur_antinode}
\end{equation}
which is again parallel to the velocity and has a quadratic nonlinear
correction.  However, the magnitude of the nonlinear term is reduced by
$1/\sqrt{2}$. This anisotropy is due to the relative positions of the
nodal lines and is insensitive to the details of the anisotropy of the
the Fermi surface or Fermi velocity because the quasiparticle states
that contribute to the current, for either orientation of the velocity,
are located in a narrow wedge, $\vartheta\le\vartheta_c\sim(v_s/v_o)\ll
1$, near the nodal lines. Thus, the occupied quasiparticle states near
any of the nodes have essentially the same Fermi velocity and density
of states; only the relative occupation of the states is modified by
changing the direction of the velocity.

The dependence of the supercurrent on the positions of the nodal lines
in momentum space suggests that the anisotropy can be used to
distinguish different unconventional gaps with nodes located in
different directions in momentum space. For example, the $d_{xy}$ state
(B$_{2g}$ representation), $\Delta\sim \hat{p}_x\ \hat{p}_y$, would
also exhibit a four-fold anisotropy, but the nodal lines are rotated by
$\pi/4$ relative to those of the $d_{x^2-y^2}$ state. The order
parameter $\Delta\sim \hat{p}_x\ \hat{p}_y(\hat{p}_x^2-\hat{p}_y^2)$,
corresponding to the $A_{2g}$ representation, would exhibit a more
complicated $2\times 4-$fold anisotropy.

Anisotropy in the in-plane current implies a similar anisotropy in the
field dependence of the in-plane penetration length.  Consider the
geometry in which the superconductor occupies the half-space $z>0$,
with $z||\hat{c}$. For a surface field $\vec{H}$ directed along a nodal
line, Maxwell's equation combined with eq. (\ref{cur_node}) for the
current and the gauge condition $\vec{\nabla}\cdot\vec{v}_s=0$
reduces to
\begin{equation}
-{\partial^2v_s\over\partial z^2}=
{4\pi e^2\over c^2}\ j_s[\vec{v}_s]=
-{v_s\over\lambda_{||}^2}\
\left\{
1-{|v_s|\over v_0}
\right\}
\,.
\end{equation}
We define the effective penetration length in
terms of the static surface impedance,
${1/\lambda_{||}}=-(1/H)(\partial b/\partial z)|_{z=0}$. The solutions for both
half-space and thin film geometries are
discussed in appendix B. We obtain,
\begin{equation}\label{lambda_node}
{1\over\lambda_{||}(\vec{H})}=
{1\over\lambda_{||}}\left(1-\frac{H}{H_0}\right)
\,,\quad\vec{H}||\,{\rm node}
\,,
\end{equation}
and similarly for fields directed along an antinode,
\begin{equation}\label{lambda_antinode}
{1\over\lambda_{||}(\vec{H})} =
{1\over\lambda_{||}}\left(1-\frac{1}{\sqrt{2}}\,\frac{H}{H_0}\right)
\,,\quad\vec{H}||\ {\rm antinode}
\,,
\end{equation}
where $\lambda_{||}$ is the zero-field penetration depth at $T=0$ and
$H_0=3c{v_0}/2e{\lambda_{||}}$ is the
characteristic field scale. Using $v_0=\mu\Delta_0/v_f^*$,
$1/\lambda_{||}^2=4\pi e^2 N_f \bar{v}_f^2/c^2$,
$\xi_{||}=\bar{v}_f/2\pi T_c$ and $H_c^2/8\pi=\frac{1}{2} N_f\Delta^2$
gives $H_0\simeq\frac{3}{2}\mu(\bar{v}_f/v_f^*)^3\,H_c\sim{\cal
O}(H_{c})$, where $H_c$ is the thermodynamic critial field, $\bar{v}_f$
is the {\it rms} average of the Fermi velocity and $v_f^*$ is the Fermi
velocity at the node. We estimate $H_c(0)\simeq 8.5\,kG$ and
$H_{0}\simeq 3.4\,T$ if we neglect the anisotropy of the Fermi
velocity. Fermi surface anistropy can change the characteristic field
significantly. Using a next-nearest neighbor tight-binding model for
the Fermi surface fit to the LDA result for
YBa$_{2}$Cu$_{3}$O$_{7-\delta}$,\cite{rad92} we estimate the anisotropy
to be $v_f^*/\bar{v}_f\simeq 1.1$, and gives $H_0\simeq 2.5\,T$;
however, the anisotropy of $\vec{v}_f$ is sensitive to the
hole concentration near half filling. Assuming $H_0=2.5\,T$ and
$H_{c1}(0)\sim 250\,G$ appropriate for twinned single crystals of
YBCO,\cite{ume90} then over the field range, $0\le H\le H_{c1}$, the
change in $\lambda_{||}(\vec{H})$ is of order
$\delta\lambda_{||}\simeq\lambda_{||}\,H_{c1}/H_{0}
\simeq\,(1,500\,\AA)(250\,G)/(2.5\times 10^4\,G)\simeq 15\,\AA$. The
magnitude of $\delta\lambda_{||}(\vec{H})$ also depends on the
measurement technique. If $\delta\lambda_{||}(\vec{H})$ is measured
from the inductive response of a low-frequency {\it a.c.} field,
$\vec{A}_{\omega}$, in the presence of a parallel {\it d.c.} field,
$\vec{v}_s$, then from eq. (\ref{cur_node})
$\delta\vec{j}_{\omega}=-\frac{e^2}{c}\,\rho
\left\{1-2\,{|\vec{v}_s|}/{v_0}\right\}\,\vec{A}_{\omega}$; and the
change in the penetration depth with the static field is a factor of
$2$ larger than the {\it d.c.} result in eqs. (\ref{lambda_node}) and
(\ref{lambda_antinode}).

Observation of the anisotropy and linear field dependence of
$\delta\lambda_{||}(\vec{H})$ could provide strong support for a
$d_{x^2-y^2}$ order parameter. Below we consider thermal and impurity
effects which might mask or wash-out the characteristic anistropy and
linear field dependence characteristic of pure material at $T=0$.

\subsubsection*{Thermal excitations: cross-over to analytic behavior}

At finite temperatures thermally excited quasiparticles occupy the
upper branch of the counter-moving band shown in Fig. 10, and for
$T\ll\Delta_0$ these thermal quasiparticles are predominantly in the
nodal regions. In the limit $v_fv_s\ll\pi T$ the thermal excitations
dominate and the nonlinear corrections can be obtained from a Taylor
expansion in $(v_fv_s/\pi T)$. In the opposite limit $\pi T\ll
v_fv_s\ll\pi T_c$, the non-thermal jets that give rise to the
non-analytic backflow current dominate. Thus, at finite temperature
there is a cross-over from the non-analytic result with
$\vec{j}_{qp}\sim e\rho\vec{v}_s\left|{v_fv_s}/{\Delta_0}\right|$
for $\pi T\ll v_fv_s\ll\pi T_c$ to $\vec{j}_{qp}\sim
e\rho\vec{v}_s\left({v_fv_s}/{\pi T}\right)^2$ for $0< v_fv_s\ll\pi
T$ (These expansions are discussed in appendix A).  Hence, with
decreasing surface field, $H$, the effective penetration depth also
crosses over from a linear field dependence,
$\delta\lambda_{||}\simeq\lambda_{||}\,(H/H_0)$, for $0<H_{T}<H\ll
H_c$, to a quadratic dependence, $\delta\lambda_{||}\sim(H/H_{T})^2$
for $0<H<H_{T}$, where the cross-over scale is the field at which the
flow energy per excitation becomes comparable to the thermal energy,
$v_fv_s\simeq v_f\frac{e}{c}H_{T}\lambda\simeq\pi T$, or
$H_{T}\simeq(T/\Delta_0)H_0$. For $T/\Delta_0=0.001 ({\it i.e.\ } T\sim
0.2\,K$ and $T_c=100\,K$) the cross-over field is $H_{T}\simeq 10\,G$
with $H_0=10\,kG$. However, at $T=2\,K$, the cross-over moves to
$H_{T}\simeq 100\,G$, which is a substantial fraction of $H_{c1}\sim
200\,G$ for the cuprates. At temperatures above $T\simeq 2\,K$ the {\it
linear} field region is essentially washed out by the thermal backflow
current. Thus, it is essential to work at $T\ll T_c(H_{c1}/H_0)$ in
order to minimize the current from the thermally excited
quasiparticles. Note that the restrictions  on the temperature are more
severe for observing the linear field dependence, compared to the
linear temperature dependence at $H=0$, because a clean interpretation
of the Meissner penetration length requires fields below the vortex
nucleation field. If vortex nucleation could otherwise be suppressed,
then the field range could be extended to $H\sim H_0$ and a linear
field dependence would be observable over a much larger field range,
and correspondingly the restriction on the low-field cross-over would
be much less severe. In any event we typically assume a field range up
to $H_{c1}\simeq 250\,G$ in our calculations and estimates.

In Fig. 13 we show the effect of thermal excitations on the velocity
dependence of the effective superfluid density $\rho_s(T,v_s)\equiv
j_s/v_s$ for $\vec{v}_s$ directed along a node.\cite{xu94} The
intercept shows the usual thermal reduction in $\rho_s$ at $v_s=0$. The
cross-over from the linear field dependence at $v_fv_s\gg\pi T$ is
clearly seen, and the arrows indicate the value of the cross-over
velocity, $v_{T}=\pi T/v_f$. Figure 14 for the field-dependent Meissner
penetration depth contains similar information. The $T=0$ result is our
analytic solution to the nonlinear London equation; for the curves at
$T\ne 0$ we converted the current-velocity relations with the same
field scaling as we derived for the $T=0$ scale, $v_s/v_0\rightarrow
H/H_0$. Stojkovi{\'c} and Valls have recently solved the nonlinear
London equations numerically at $T\ne 0$.\cite{sto94} Our scaling
assumption for $T\ne 0$ agrees well with their numerical solutions for
$H_{T}\le H_{c1}$. Figure 14 gives the estimates of
$\delta\lambda_{||}$ for $H\simeq H_{c1}$ for several temperatures.
The two estimates at $T=0$ correspond to $H_{c1}/H_{0}(d.c.)=0.01$,
appropriate for the {\it d.c.} measurement of $\delta\lambda_{||}$, and
$H_{c1}/H_{0}(a.c.)=0.02$, appropriate for the {\it a.c.} measurement
of $\delta\lambda_{||}$. Thus, at $T\rightarrow 0$ we estimate a change
in $\lambda_{||}\sim 30\,\AA$ in an {\it a.c.} measurement for
fields directed along the node. At $T=0.4\, K$ the cross-over is
observable in the calculation, but we expect to see a clear linear
field dependence over the field range up to $H_{c1}$. At $T\simeq 1\,K$
the resolution is reduced, $\delta\lambda_{||}\simeq 20\,\AA$ and a
linear field dependence is observed over roughly $60\,\%$ of the field
range. However, by $T\simeq 4\,K$ the cross-over field is $H_{T}\simeq
H_{c1}$; the linear field dependence is washed out and the change in
$\lambda_{||}$ is less than $15\,\AA$. These results are all
reduced by a factor of $\simeq 1/\sqrt{2}$ for a field along an
antinode.

\begin{figure}
\centerline{\psfig{file=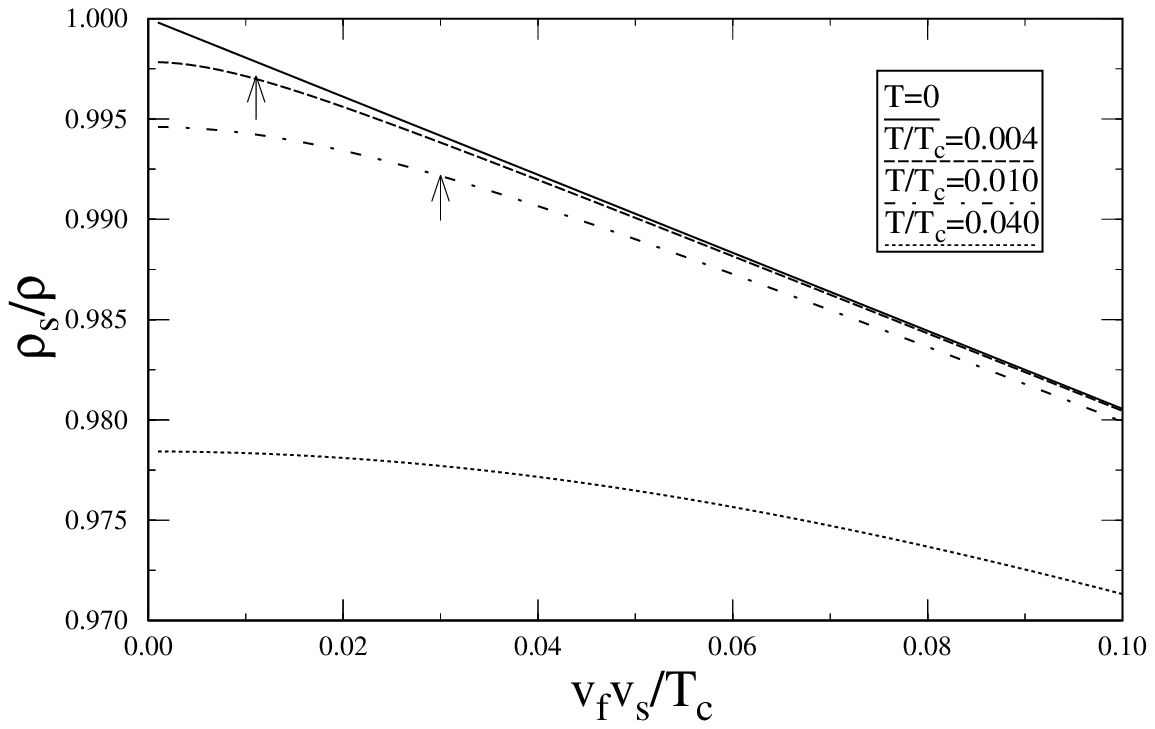,height=3in}}
\begin{quote}
\small
Fig. 13 {\hskip 10pt} Velocity and temperature dependence of the
effective superfluid density, $\rho_s=j_s/v_s$, for $\vec{v}_s ||\,{\rm
node}$ in the clean limit. The cross-over velocities are indicated by the
arrows.
\end{quote}
\end{figure}

\begin{figure}
\centerline{\psfig{file=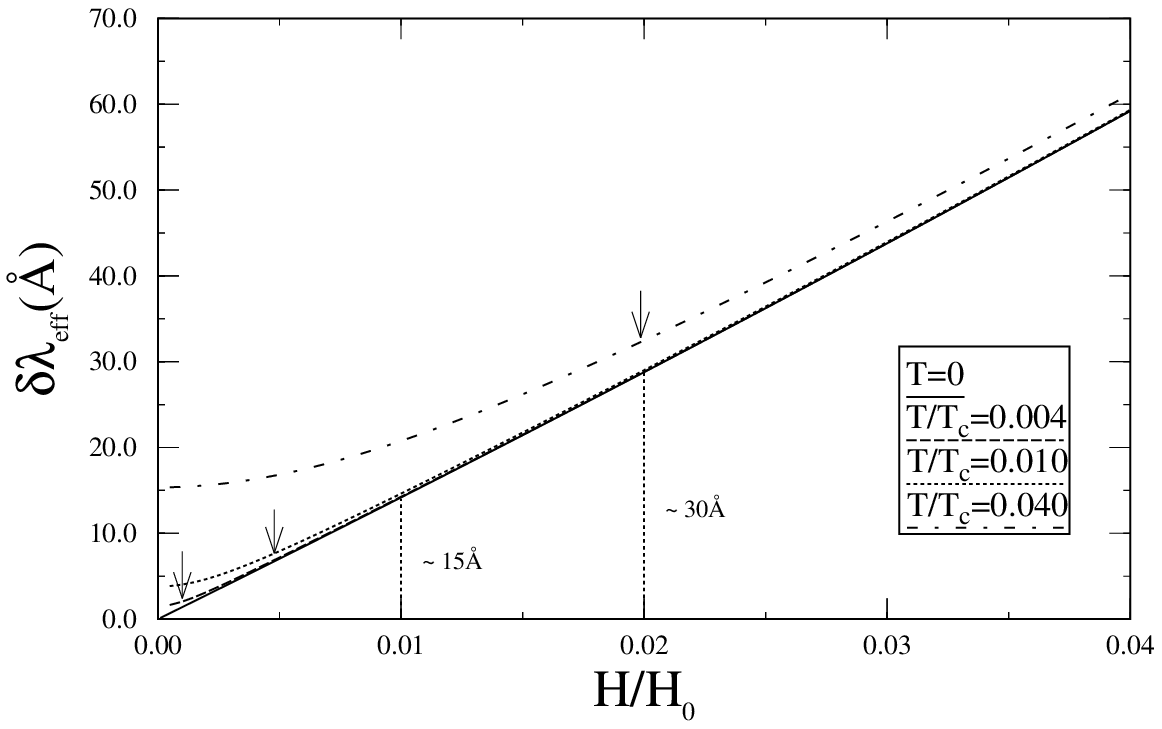,height=3in}}
\begin{quote}
\small
Fig. 14 {\hskip 10pt} Field dependence of the penetration depth for
$\vec{H}||\,{\rm node}$. The cross-over field, $H_{T}=(T/\Delta_0)H_0$,
is indicated by an arrow. The verticle marker at $0.01$ ($0.02$) corresponds to
$H_{c1}/H_{0}$ for a {\it d.c.} ({\it a.c.}) measurement of the penetration
depth.
\end{quote}
\end{figure}

\subsubsection*{Impurity scattering}

Quasiparticle scattering by impurities also removes the non-analyitc
dependence of the current on the condensate flow velocity at
sufficiently low $v_s$. At $T=0$ impurity scattering gives rise to a
cross-over velocity $v_s^*$, or field $H^*=H_0(v_s^*/v_0)$, that is
determined by the energy scale, $\varepsilon^*$, in the density of
states $\bar{{\cal N}}(E)$; above $\varepsilon^*\ll v_fv_s\ll \pi T_c$
the excitations that are strongly affected by impurity scattering at
energies $E\lesssim\varepsilon^*$ are only a small fraction of the
non-thermal backflow current, while at small flow velocities,
$v_fv_s\lesssim\varepsilon^*$ the excitations contributing to the
non-thermal backflow current are strongly modified by impurity
scattering.

The cross-over field scale at $T=0$ due to impurity scattering can be
obtained from the general expression in eq. (\ref{current_mats}) for
the current; the cross-over velocity is given by
$v_fv_s^*=\varepsilon^*$, where $\varepsilon^*$ is the cross-over energy from
eq. (\ref{cross_over_scale}).
In the Born limit ($\delta_0\ll 1$) the cross-over scale,
\begin{equation}\label{impurity_crossover}
H^*=H_0\,\frac{\varepsilon^*}{\mu\Delta_0}\simeq
\frac{2}{\mu} H_0\,e^{-\frac{\mu}{2}\pi\tau\Delta_0} =
\frac{2}{\mu} H_0\,e^{-\frac{\mu}{2}(l_{imp}/\xi_0)}
\quad\quad {\rm (Born)}\,,
\end{equation}
is exponentially small for $l_{imp}\gg \xi_0$. Thus, the linear field
dependence and anisotropy of the penetration depth will be unaffected
by impurity scattering in the weak scattering limit.  The cross-over
field may be much higher, even in the dilute limit, for strong
scattering. In the unitarity limit ($\delta\rightarrow \pi/2$) the
cross-over field scale is
\begin{equation}\label{unitarity_crossover}
H^*\simeq
H_0\,\sqrt{\frac{\pi}{2\ln(\Delta_0/\Gamma_u)}\,\frac{\Gamma_u}{\mu\Delta_0}}
\quad\quad {\rm (Unitarity)}
\,.
\end{equation}
For $\Gamma_u/\Delta_0=10^{-4}$, which is a good bound for $\Gamma_u$
obtained from the data of Ref. (\onlinecite{har93}), we obtain,
$H^*\simeq 2.6\times 10^{-3}\,H_0\simeq 26\,G$. Thus, even in the
unitarity limit there is a large field range, $25\,G\sim H^* < H<
H_{c1}\sim 250\, G$ in which the linear field dependence of
$\delta\lambda_{||}(\vec{H})$ is expected to hold. Of course it is
important to be at low temperatures in order to avoid the thermal
cross-over. Figure 15 shows our numerical results for the
field-dependent penetration depth ({\it d.c.}) calculated for
$T/T_c=0.004$ and unitarity scattering. The field range is
approximately $0<H\lesssim H_{c1}\simeq 250\,G$. In the clean limit
($\Gamma_u=0$) the curvature at very low fields is due to the thermal
cross-over discussed earlier.  As the impurity concentration increases
the curvature sets in at higher fields; the arrows indicate the
impurity cross-over fields calculated from eq.
(\ref{unitarity_crossover}); note the calculated cross-over field
accurately reflects the field dependence obtained from the numerical
calculations. The results for $\Gamma_u/\Delta_0\simeq 10^{-4}$
suggest that a nonlinear Meissner effect with a linear field
dependence over $\sim 80\,\%$ of the field range from zero to $H_{c1}$
should be observable in single crystals of comparable purity to those
of Ref. (\onlinecite{har93}).

\begin{figure}
\centerline{\psfig{file=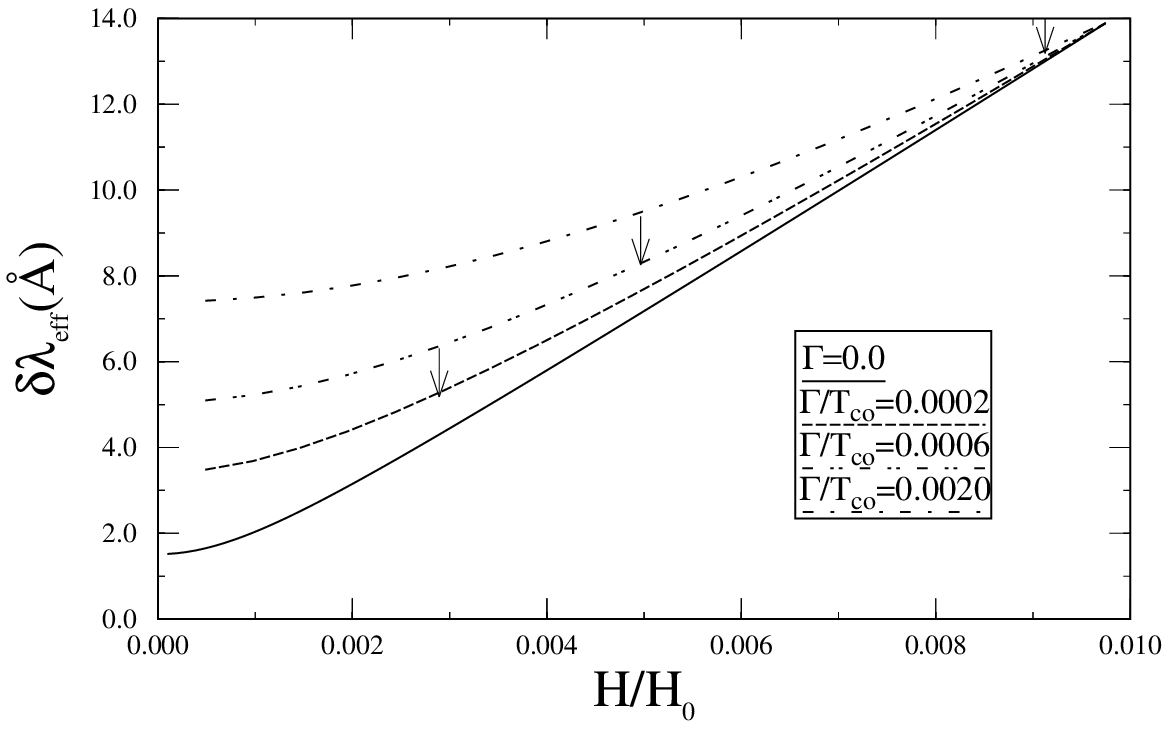,height=3in}}
\begin{quote}
\small
Fig. 15 {\hskip 10pt} Field-dependence of the penetration depth with unitarity
scattering. The cross-over field $H^*$ is indicated by an arrow.
\end{quote}
\end{figure}

\section{Anisotropy of the in-plane magnetic torque}

Another test of the presence of nodal lines associated with a
$d_{x^2-y^2}$ order parameter would be to measure the magnetic
anisotropy energy, or magnetic torque, for in-plane
fields.\cite{yip92a}  Consider a velocity field
$\vec{v}_s=v_{x'}\,\hat{x}'+v_{y'}\,\hat{y}'$ that is not directed
along a node or antinode. At $T=0$ the projections of the velocity
along the nodal lines $\hat{x}'$ and $\hat{y}'$ give rise to two
backflow jets of different magnitudes,
\begin{equation}
\vec{j}_{qp}=-eN_f v_f^2
\Bigg[\
\left({v_{x'}^2\over v_o}\right)\ (-\hat{x}')+
\left({v_{y'}^2\over v_o}\right)\ (-\hat{y}')
\ \Bigg]\ .
\label{jgendir}
\end{equation}
The important feature is that the current is not parallel to the
velocity field, except for the special directions along the nodes or
antinodes. As a consequence the magnetic field in the screening layer,
$\vec{b}$, is not parallel to the applied surface field, $\vec{H}$.
This implies that there is an in-plane magnetic torque which acts to
align the nodes of the gap, and therefore the crystal axes, with the
surface field.

The magnetic free energy of the superconductor, in the presence of the
surface field $\vec{H}$, is given by
\begin{equation}
U=-\int_{V_{film}}\,d^3x\int_0^H\vec{M}(\vec{H}')\cdot d\vec{H}'
\,.
\end{equation}
The integration is carried out assuming that the orientation of
$\vec{H}'$ is fixed;
$\vec{H}'=H'(\sin(\theta)\hat{x}'-\cos(\theta)\hat{y}')$, and
$\vec{M}(\vec{H})$ is the equilibrium magnetization for the given value
of $\vec{H}$, which is easily be found by solving the nonlinear London
equation in the film geometry (see appendix B). The resulting
anisotropic contribution to the magnetic energy is obtained by
integrating over the volume of the film,\footnote{The authors of Ref.
(\cite{yip92a}) erred in calculating the magnetic anisotropy energy by a factor
of $3$. The corrected result is given by eq.(\ref{mag_anisotropy}).}
\begin{equation}\label{mag_anisotropy}
U_{an}(\theta)=
-\frac{H^2}{4\pi}\frac{H}{H_0}\,A\,\lambda_{||}\,\left<\Phi\right>
\left[\sin^3(\theta) + \cos^3(\theta)\right]
\,,
\end{equation}
where $A$ is the surface area of the film, and (see appendix B)
\begin{equation}
\left<\Phi\right>=2\int_0^{d/2\lambda}\,d\zeta\,\Phi(\zeta)
=\frac{8}{3}
\left\{
\frac
{\sinh^4(\frac{d}{4\lambda})
  \left[\frac{1}{2}+\cosh^2\left(\frac{d}{4\lambda}\right)\right]}
{\cosh^3\left(\frac{d}{2\lambda}\right)}
\right\}
\,.
\end{equation}
In the thin and thick film limits we obtain,
\begin{equation}
\left<\Phi\right>=\Bigg\{
\begin{array}{l}
\frac{1}{4}\left(\frac{d}{2\lambda}\right)^4 \quad , d\ll\lambda_{||}\\
\frac{1}{3}\qquad \qquad , d\gg \lambda_{||}\,.
\end{array}
\end{equation}
Note that the anisotropy energy is minimized for field directions along
the nodal lines, and is maximum for fields along the antinodes.

At finite temperature an analytic expression corresponding to eq.
(\ref{jgendir}) is not available except in special limits ( see
appendix A). In order to calculate the magnetic torque at finite
temperature, we note that for $ \vert v_f v_s \vert \ll \Delta_0$, we
can write $ j_{s \,x, y} = \rho_{s \,x,y} v_{s \,x, y} $ (we drop the
primes on $x$ and $y$, but emphasize that the axes refer to two
orthogonal nodal directions) with
\begin{equation}
\frac{\rho_{s \ x,y}}{\rho} =
 1-\mbox{\large $g$}\left(v_fv_{s \ x,y}/\mu\Delta_0\right)\,,
\end{equation}
where $g(x)$ is a dimensionless function (see the subsections on
`thermal excitations' and `impurity scattering' in section IV).  At
$T=0$ $g(x)=|x|$, manifesting the non-analytic behavior of the current,
while at finite temperature $g(x)$ has a linear region for large $x$
(${v_fv_s / T} \gg 1$), and crosses over to a quadratic region for
small $x$ (${v_fv_s / T }\ll 1$).

The flow velocity and field distribution are determined by the nonlinear London
equation,
\begin{equation}
{\partial^2 v_{s \ x,y} \over \partial z ^2 }
- \frac{1}{\lambda_{||}^2}\ v_{s \ x,y}
\left\{
1-\mbox{\large $g$}\left({v_fv_{s \ x,y}/\mu\Delta_0}\right)
\right\} =0
\,,
\end{equation}
where $\lambda_{||}$ is the penetration length in the limit of zero
field.  The solution for the velocity field can be obtained by
performing a perturbation expansion in $v_f v_s/\mu\Delta_0$. For a
superconductor occupying the half space $z > 0$, the solution is
\begin{equation}
u_{x,y}(z)={v_fv_{s\,x,y} \over \mu\Delta_0}
= u_{x,y\,0} e^{-z/\lambda}+ e^{-z/\lambda} \int_0^z \frac{dz'}{\lambda}
\frac{{\cal G}\left(u_{x,y\,0} e^{-z'/\lambda}\right)}{u_{x,y\,0}}
e^{2z'/\lambda}\,,
\end{equation}
where ${\cal G}(x)=\int_o^x x'g(x')dx'$. The value of the $\vec{u}_0$ is
determined by the boundary condition, ${d{v}_{s \ x,y} / dz}|_{z=0}=\pm
\frac{e}{c} {H}_{y,x}$; thus,
\begin{equation}
u_{x,y\,0} = \mp \frac{H_{y,x}}{\frac{2}{3}H_0}+\,
\frac{{\cal G}\left({H_{y,x}\over
\frac{2}{3}H_0}\right)}{{H_{y,x}\over\frac{2}{3}H_0}}
\,.
\end{equation}
The magnetic torque can be obtained from the magnetic anisotropy energy, $
\tau_z = - {\partial U_{an}}/{\partial \theta}$, or
equivalently $\vec M \times \vec H$. The integral of
$\vec{b}=\vec{\nabla}\times\vec{A}$ can be expressed in terms of $\vec{v}_s$ at
the surface, since $\int_0^{\infty} b_{x,y}  dz = \pm \frac{c}{e} {v}_{s \ y,x}
|_{z=0} $.
Thus, for a thick film
\begin{equation}\label{torquevsT}
\tau_z = 2\frac{1}{4\pi} (\frac{2}{3}H_0)^2\lambda A\,
\left[\frac{H_y}{H_x} {\cal G}\left({H_x\over \frac{2}{3}H_0}\right) +
\frac{H_x}{H_y}{\cal G}\left({H_y\over \frac{2}{3}H_0}\right)
\right]
\,,
\end{equation}
the torque can be calculated from a simple integration of the
current-velocity relation.  At $T=0$ the torque equation
(\ref{torquevsT}) reduces to
\begin{equation}\label{torque}
\tau_z=\frac{1}{4\pi}\lambda \frac{H}{H_0} H^2 A \sin\theta\cos\theta
(\sin\theta-\cos\theta)\quad , \quad 0\leq\theta\leq\pi/2
\,,
\end{equation}
in agreement with the derivative of the anisotropy energy in eq.
(\ref{mag_anisotropy}).

For $H=400\,G$, $A=(2,000\,\mu{\rm m})^2$ and $\lambda=1,400\AA$,  the
zero temperature  maximum magnetic torque $\tau_z \simeq
(1/12\sqrt{3}\pi)H^2(H/H_0)A\lambda \sim 10^{-5}\, {\rm dyne-cm/rad}$.
In Fig. 16 we show results for the magnitude of the torque at finite
temperature. The torque is only weakly reduced for $T\lesssim\,1\,K$,
but drops rapidly above $T\sim 1\,K$.

\begin{figure}
\centerline{\psfig{file=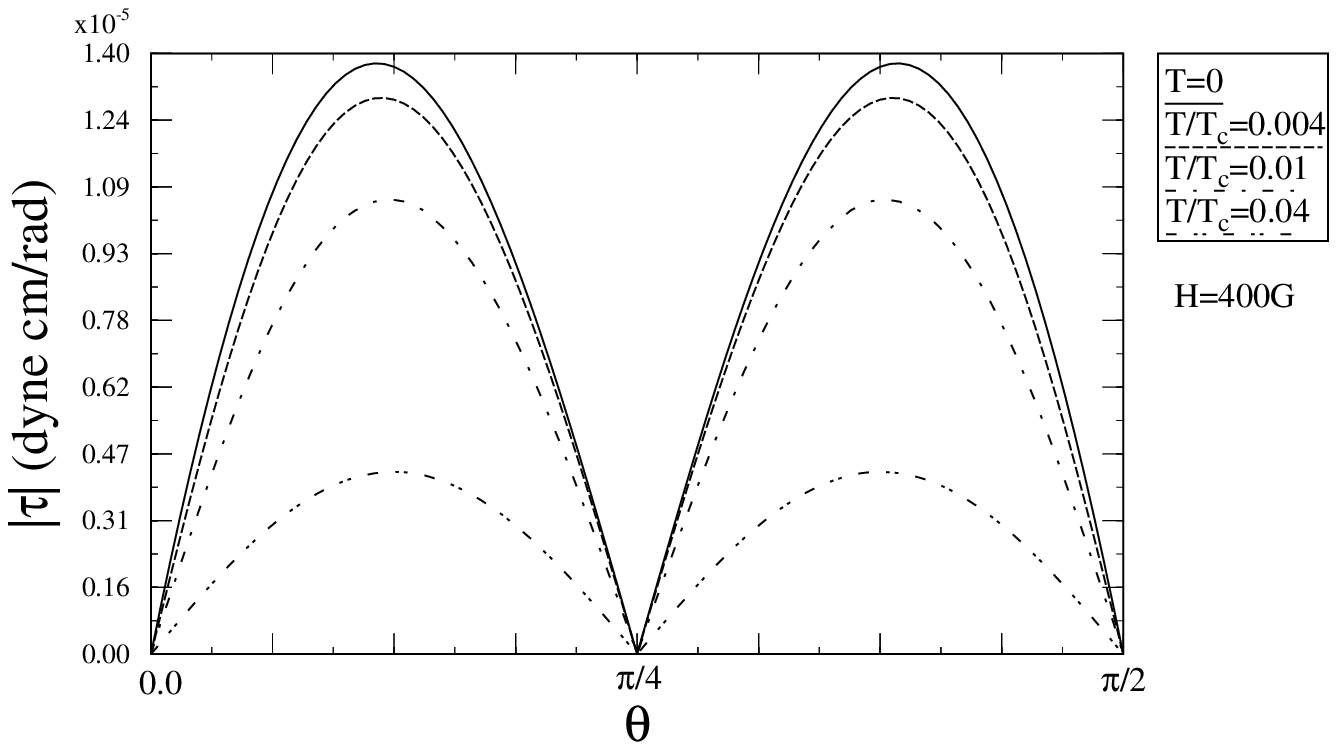,height=3in}}
\begin{quote}
\small
Fig. 16 {\hskip 10pt} Magnitude of torque as a function of $\theta$.
Note that $\theta=0,\pi/2$ correspond to node positions, $\theta=\pi/4$
corresponds to an antinode position, and the maximum torque (for $T=0$)
occurs at $\theta=\frac{1}{2}\sin^{-1}(2/3)\simeq 21^{o}$. The torque
has four fold symmetry and points to the nodal  positions.
\end{quote}
\end{figure}

In order to maximize the torque it is desirable to suppress vortex
nucleation so that the torque measurement can be performed at a higher
magnetic field. In this respect, thin films with dimensions
$d\le\lambda$ might be desirable, because the vortex nucleation field is
increased by roughly $(\lambda/d)$ in a thin film. The optimum geometry
might be a superlattice of superconducting/normal layers with an
S-layer thickness, $\xi\ll d<\lambda$. In this case the field at each
SN interface is essentially the external field, and the anisotropy
energy is enhanced by the number of S-layers.

\subsubsection*{Measurements of the transverse magnetization}

In a recent paper Buan, {\it et al.} reported\cite{bua94} measurements
of the transverse magnetic moment induced by an in-plane surface field
in an untwinned single-crystal of LuBa$_2$Cu$_3$O$_{7-\delta}$. The
surface field was rotated in the a-b plane and the Fourier component
of $M_{\perp}(\theta)\propto\cos(4\theta)$ was extracted and compared
with predictions based on the theory of Ref. (\cite{yip92a}). Buan, {\it
et al.} report a measured transverse magnetization signal
($\sim\cos(4\theta)$) of $M_{\perp}^{exp}\simeq 0.8\times 10^{-6}\,
emu$ at $H=300\,G$ and $T=2\,K$, and a resolution limit of
$0.3\times 10^{-6}\,emu$. Buan, {\it et al.} also report a theoretical value of
$M_{\perp}^{theory}\simeq 2\times 10^{-6}\, emu$, for the same temperature and
field, obtained from numerical solution of the London
equation with the nonlinear current-velocity equation from Ref.
(\cite{yip92a}). This estimate is 2.5 times the experimental signal and nearly
an order of magnitude above the resolution limit.
The authors conclude that the $\cos(4\theta)$ signal is too small to be
consistent with a pure d-wave pairing state, that there are no nodes in
the gap and that the $\cos(4\theta)$ signal is consistent with a higher
harmonic of the $\cos(\theta)$ and $\cos(2\theta)$ signals associated
with the shape anisotropy of the crystal.

While it may be that the observed $\cos(4\theta)$ signal is associated
with extrinsic effects of geometry, the conclusion that the measurement
rules out a pure d-wave state with nodes relies principally on
$M_{\perp}^{theory}\gg M_{\perp}^{exp}$.  The estimate of
$M_{\perp}^{theory}\simeq 2\times 10^{-6}\,emu$ at $T=2\,K$ and
$H=300\,G$ is based on the following parameters:  $\lambda_c =1\,\mu
m$, $\lambda_{||}=1,700\,\AA$ and geometric parameters for the crystal,
$a=1.2\, mm$, $b=0.9\, mm$ and $c=0.07\, mm$.  The transverse magnetic
moment per unit area is given by $M_{\perp}=\tau_z/(Area\, H)$, which
is proportional to $H^2\lambda_{||}^2$ from eq.
(\ref{torque}) for $T\rightarrow 0$. Buan, {{\it et al.\ }} argue that finite
size
effects require that $\lambda_{||}$ in the formula for $M_{\perp}$ be
replaced by an effective penetration depth $\lambda\simeq 4,000\,\AA$
in order to account for c-axis currents. This procedure leads to an
{\it increase} in the in-plane transverse magnetization due to current
flow along the c-axis, which is opposite to what is expected. In
particular, in a geometry with an aspect ratio, $c/a\simeq 0.07$, and a
field lying in the ab-plane, say along the $b$-axis, the current flows
predominantly along the $a$-axis. The `return current' at the edges
flows mainly along the c-axis. The main effect of the c-axis
currents is a reduction in the area with current flow in the a-b plane,
$A_{eff}\simeq A(1-2\lambda_c/a)$. Thus, the in-plane,
transverse magnetization for the semi-infinite geometry will
correspondingly be {\it reduced} by roughly $A_{eff}/A =
(1-2\lambda_c/a)$. The reduction factor is tiny for the geometry of
Ref. (\cite{bua94}), $A_{eff}/A \simeq 0.999$; however, the theoretical
value for $M_{\perp}^{theory}$ reported in Ref.  (\cite{bua94}) is
overestimated by a factor of $(\lambda/\lambda_{||})^2\simeq 5.5$.
Dividing the theoretical estimate of $2\times 10^{-6}\,emu$ by $5.5$
gives $M_{\perp}^{theory}\simeq 0.36\times 10^{-6}\,emu$, very near the
resolution limit and below the observed signal at $H=300\,G$.

Thus, in our opinion this null result is inconclusive and does not
force one to eliminate a pure d-wave state as a possible candidate for
the order parameter of the cuprates. Experiments designed to minimize
shape anisotropies at temperatures well below $T\simeq 2\, K$ should be
able to detect the intrinsic anisotropy associated with nodal
excitations, should they exist.

\section{Acknowledgements}

We wish to thank Walter Hardy, Doug Bonn and their colleagues for
sending us their results prior to publication. After completing this
work, but before this manuscript was complete, we received a preprint
from Stojkovi\'c and Valls reporting numerical calculations of the
field dependent penetration depth and transverse magnetization at
finite temperatures. We thank the authors and note that our results are
in good agreement when allowances are made for different choices of
parameters (SV generally examined a much larger range of fields) and
different details for the model of the gap.  This work was supported by
the National Science Foundation (DMR 91-20000) through the Science and
Technology Center for Superconductivity.

\newpage
\section*{Appendix A - Temperature and Field Dependence of the Current for a
Superconductor with Line Nodes}

In general we need to solve two problems, (i) the self-consistent
calculation of the order parameter, $\Delta(\vec{p}_f)$, in the
presence of a condensate flow field at finite temperature, and (ii) the
computation of the temperature and field dependence of the current,
$\vec{j}_s$, for a given $\Delta(\vec{p}_f; \vec{v}_s, T)$. These
problems are in general coupled because the flow field leads to
suppression of the gap parameter, $\Delta$, which contributes to the
field-dependence of the current. We show that at low temperatures,
${T}/{T_c} \ll 1$, and low fields, ${v_f v_s}/{T_c} \ll 1$, the
contributions to the current from the gap suppression can be
neglected.

We also discuss the functional form of the current at low temperatures
and low velocity in the clean limit. In the limit ${T/ T_c} \ll 1$
and ${v_f v_s / T} \ll 1$, an expansion in $v_f v_s$ is valid, and
as a result, the leading field dependence of the superfluid density,
$\rho_s$, is quadratic in the flow field; while in the limit ${T/
T_c} \ll 1$ and ${v_f v_s / T} \gg 1$, the quasiparticle
contribution to the current is non-analytic. For simplicity we confine
ourselves to cases where $\vec v_s$ is parallel to a node or an
antinode, in which case $\vec j_s$ and $\vec v_s$ are parallel so we
drop the vector symbols.

In the limit $\vert v_f v_s \vert  \ll T \ll \Delta_0$ we can expand
$j_s$ in a power series in $v_s v_f$,

\begin{equation}
j_s = - e N_f v_f^2 \ {v}_s
    \left[ 1 - \gamma_1 \left({T \over \Delta_0}\right) -
         \gamma_2 {(v_f v_s)^2 \over T \Delta_0} -  ... \ \right]
\label{a1}
\end{equation}
where $\Delta_0 = \Delta_0 (v_s, T)$, and $\gamma_1, \gamma_2$ are
numerical coefficients of order one whose values depends on the angular
slope parameter $\mu$ defined in eq.(\ref{dx2-y2_gap}). The $\gamma_1$
term is the usual linear dependence of $\rho_s$ on $T$. The $\gamma_2$
term, being proportional to $1/T$, signals the breakdown of the Taylor
expansion in $v_s$ for sufficiently low temperatures.

The magnitude of the gap has the expansion,
\begin{equation}
\Delta_0(v_s,T) = \Delta_0(0,0)
\left[1 - \alpha_1 \left({ T \over \Delta_0(0,0) } \right)^3
        -  \alpha_2 \left({v_f v_s \over \Delta_0(0,0) }\right)^2
             \left({ T \over \Delta_0(0,0) }\right)  - ... \,\right]
\label{a2}
\,.
\end{equation}
Thus, it is clear that the leading correction to $j_s$ from the flow
field $v_s$ is given entirely by the $\gamma_2$ term; gap
suppression is a higher-order correction.

In the limit $ T \ll \vert v_f v_s \vert \ll \Delta_0$, we obtain,
\begin{equation}
j_s =  -e N_f v_f^2 \ {v}_s  \
\left[1 - \gamma'_1 {\vert v_f v_s \vert \over \Delta_0}-
\gamma'_2\left({T^2 \over \vert v_f v_s \vert \Delta_0}\right)-  ... \,
\right]
\label{a3}
\,.
\end{equation}
The $\gamma'_1$ term was calculated earlier, and the $\gamma'_2$ term
can be obtained from the Sommerfeld expansion of integrals involving
the Fermi function, with $\vert v_f v_s \vert$ playing the role of the
chemical potential for the quasiparticles. Note that both coefficents
depend on the direction of $\vec{v}_s$. Also not that the $\gamma'_2$ term
gives a non-analytic current independent of $ \vert v_s \vert $, but
the result is only valid for  $ T \ll \vert v_f v_s \vert $. The gap in the
above formula is
\begin{equation}
\Delta_0(v_s,T) = \Delta_0(0,0)
\left[ 1 - \alpha'_1  ({\vert v_f v_s \vert \over \Delta_0(0,0)} )^3
   - \alpha'_2\left({\vert v_f v_s \vert \over \Delta_0(0,0)}\right)
            \left({ T \over \Delta_0(0,0)}\right)^2  - ... \,\right]
\label{a4}
\,,
\end{equation}
which gives higher-order corrections to $j_s$. The quadratic correction
to $j_s$ is unaffected by the gap suppression; the lowest-order
temperature correction is given by the term $\propto \gamma'_2 \left({T^2
\over \vert v_f v_s \vert \Delta_0}\right)$.

\section*{Appendix B - Nonlinear London Equations}

\subsubsection*{Conventional superconductor with an isotropic gap}

We include solutions of the nonlinear London equations for a parallel
surface magnetic field penetrating into a superconductor. Starting from
Ampere's equation
\begin{equation}
\vec{\nabla}\times(\vec{\nabla}\times\vec{A}) = \frac{4\pi}{c}\vec{j}_s
\,,
\end{equation}
consider first an isotropic superconductor with a conventional gap. The
constitutive equation for the current is given by eq.(\ref{current_swave}).
In the absence of phase vortices, $\vec{\nabla}\times\vec{\nabla}\chi=0$, we
obtain the nonlinear London equation,
\begin{equation}
\nabla^2\vec{v}_s - \frac{1}{\lambda^2}\,\vec{v}_s
\left\{
1-\alpha(T)\left(\frac{v_s}{v_c}\right)^2
\right\} =0
\,,
\end{equation}
where we have chosen the gauge, $\vec{\nabla}\cdot\vec{v}_s=0$. This
gauge choice is consistent with current conservation provided that the
current is transverse to the spatial variation of the field, which
requires that there be no vortices present. Consider the geometry in
which the superconductor occupies the half-space $z>0$. A surface field
$\vec{H}=H\hat{x}$ is parallel to the interface, and the screening
current $\vec{v}_s$ is then parallel to $\hat{y}$. The field in the
superconductor is given by $\vec{b} =
\frac{c}{e}({dv_s}/{dz})\hat{x}$. Continuity of the parallel field at
the interface imposes the boundary condition
\begin{equation}
\frac{dv_s}{dz}\vert_{z=0} = \frac{e}{c}H
\,.
\end{equation}
The field, and therefore the screening current, also vanish deep inside
the superconductor. We introduce a dimensionless velocity,
$u=\sqrt{\alpha}v_s/v_c$, and distance, $\zeta = z/\lambda(T)$; the
differential equation for $u(\zeta)$ becomes,
\begin{equation}
\frac{d^2u}{d\zeta^2}-u(1-u^2)=0
\,,
\end{equation}
with the boundary conditions,
\begin{equation}\label{bc}
\frac{du}{d\zeta}\vert_{\zeta=0} =
\frac{e}{c}\frac{\sqrt{\alpha}v_f H \lambda}{\Delta}\equiv h_0
\qquad , \qquad
u({\zeta\rightarrow\infty}) = 0
\,.
\end{equation}
A first integral is obtained by multiplying by ${du}/{d\zeta}$, integrating,
and applying the asymptotic boundary condition, $u\rightarrow 0$ as
$z\rightarrow\infty$, to obtain,
\begin{equation}
\frac{du}{d\zeta}=-u\sqrt{1-\frac{1}{2}u^2}
\,.
\end{equation}
In the physically relevant limit, $|u|\ll 1$, we replace
$({1-\case{1}{2}u^2})^{1/2}\rightarrow(1-\case{1}{4}u^2)$, and
integrate to obtain,
\begin{equation}
u=\frac{u_0}{\sqrt{u_0^2/4 + (1-u_0^2/4)e^{2\zeta}}}
\,.
\end{equation}
The dimensionless magnetic field is given by
\begin{equation}
h(\zeta)=\frac{du}{d\zeta}=-\frac{u_0(1-u_0^2/4)\,e^{2\zeta}}
{\left[u_0^2/4 + (1-u_0^2/4)e^{2\zeta}\right]^{3/2}}
\,,
\end{equation}
and the constant $u_0$ is determined by the boundary condition on magnetic
field at the interface, eq.(\ref{bc}); $h_0=-u_0(1-u_0^2/4)$, or for weak
nonlinearity,
$u_0\simeq -h_0(1+h_0^2/4)$,
which yields the solution,
\begin{equation}
h=\frac{h_0\,e^{2\zeta}}
{\left[h_0^2/4 + (1-h_0^2/4)e^{2\zeta}\right]^{3/2}}
\,.
\end{equation}
Note that if we neglect the $h_0^2$ terms in the denominator we recover the
linear response solution to the London equation, $h=h_0\,e^{-\zeta}$.

A strong surface field produces a reduction in the screening current at
the surface and a correspondingly longer initial penetration depth. We
take the initial decay rate to define the effective penetration depth,
\begin{equation}
\frac{1}{\lambda_{eff}}(h_0)\equiv\frac{1}{\lambda}
\left[-\frac{1}{h_0}\frac{dh}{d\zeta}\right]
=\frac{1}{\lambda}
\left[1-\frac{3}{4}h_0^2\right]
\,.
\end{equation}
This definition is equivalent to identifying the effective penetration
depth with the surface impedance, {\it i.e.\ }
${1}/{\lambda_{eff}}(h_0)\propto{j_s(0)}/{H}$. Note that the
effective penetration depth increases with field as expected.
In physical units
\begin{equation}
h_0=\lambda\sqrt{\alpha}\frac{v_f}{\Delta}\frac{e}{c} H
   =\sqrt{\alpha}\frac{H}{H_0}
\,,
\end{equation}
where $H_0={c\Delta}/{e\lambda v_f}
\simeq{\phi_0}/{\lambda\xi}\simeq H_c$ and $H_c$ is the
thermodynamic critial field.

\subsection*{$d_{x^2-y^2}$ superconductors}

Now consider a superconductor with a $d_{x^2-y^2}$ order parameter at
$T=0$. Choose a coordinate system in which the nodes of the gap are
directed along the $\pm\hat{x}$ and $\pm\hat{y}$ axes. The nonlinear
London equations for the corresponding projections of the condensate
velocity are
\begin{equation}
\frac{d^2u_i}{d\zeta^2} - u_i
\left\{
1-u_i
\right\} =0
\,\qquad i=x,y
\,,
\end{equation}
where $u_i = v_i/v_0$, $\zeta=z/\lambda_{||}$, and the velocity and
length scales are $v_0={\mu\Delta_0}/{v_f}$ and
$\lambda=\lambda_{||}$. These equations
are to be solved subject to the boundary condition
$\vec{H}=\vec{\nabla}\times\vec{A}|_{z=0}$, which becomes,
\begin{equation}
\frac{du_x}{d\zeta}\vert_{\zeta=0}=-h_0\,\cos\theta
\qquad , \qquad
\frac{du_y}{d\zeta}\vert_{\zeta=0}=-h_0\,\sin\theta
\,,
\end{equation}
where $\theta$ is the angle of $\vec{H}$ measured relative to the node along
$-\hat{y}$, and $h_0=\frac{e}{c}\lambda H/v_0$. Note that $h_0\sim H/H_0$. The
differential equation can be solved perturbatively; the first integral is
\begin{equation}
\frac{du_i}{d\zeta}=-u_i\left(1-\frac{2}{3}u_i\right)^{1/2}
\,,
\end{equation}
which can be integrated to give
\begin{equation}
u_i(\zeta)=\frac{u_{i0}}
{\frac{1}{3}u_{i0} +
\left[
(1-\frac{1}{3}u_{i0})\cosh(\zeta)+\sqrt{1-\frac{2}{3}u_{i0}}\sinh(\zeta))
\right]}
\,,
\end{equation}
where $u_{i0}$ is the value of the velocity field at $\zeta=0$, which is fixed
by the boundary conditions on the field. In the limit $h_0\ll 1$ we obtain
\begin{equation}
u_{x0}=h_0\,\cos\theta\left(1+\frac{1}{3}h_0\,\cos\theta\right)
\,,
\end{equation}
\begin{equation}
u_{y0}=h_0\,\sin\theta\left(1+\frac{1}{3}h_0\,\sin\theta\right)
\,.
\end{equation}
The magnetic field in the screening layer,
$\vec{b}=\vec{\nabla}\times\vec{A}$, is given by
\begin{equation}
b_x/H = -\frac{1}{h_0}\left(\frac{du_y}{d\zeta}\right)\simeq
+\sin\theta\,e^{-\zeta}\left(1+\frac{2}{3}h_0\sin\theta[1-e^{-\zeta}]\right)
\,,
\end{equation}
\begin{equation}
b_y/H = +\frac{1}{h_0}\left(\frac{du_x}{d\zeta}\right)\simeq
-\cos\theta\,e^{-\zeta}\left(1+\frac{2}{3}h_0\cos\theta[1-e^{-\zeta}]\right)
\,.
\end{equation}
Note that the field in the screening layer is not parallel to the applied
surface field. This leads to an in-plane magnetic torque acting on the
superconductor which tends to align the nodes of the order parameter and the
surface field.

\subsection*{Field penetration in a thin film $d_{x^2-y^2}$ superconductor}

A geometry in which the torque anisotropy may be measured is a film
of thickness $d\sim\lambda_{||}$. The field $\vec{H}$ is oriented as in
the half-space geometry, parallel to the $CuO$ planes and the surfaces
of the film. Choose the origin at the center of the film; the boundary
conditions for both interfaces are now,
\begin{equation}
\frac{du_x}{d\zeta}\vert_{\pm d/2\lambda}=-h_0\,\cos\theta
\qquad ,\qquad
\frac{du_y}{d\zeta}\vert_{\pm d/2\lambda}=-h_0\,\sin\theta
\,.
\end{equation}
We solve the differential equation perturbatively. Let
$u_i(\zeta)=L_i(\zeta)+\alpha_i(\zeta)$, where $L_i(\zeta)$ is the
solution to the boundary-value problem for the linearized differential
equations, {\it i.e.\ }
\begin{equation}
L_i(\zeta)= a_i\,\sinh(\zeta)
\,,
\end{equation}
with $a_i$ fixed by the boundary conditions,
\begin{equation}
a_x\cosh(\frac{d}{2\lambda})=-h_0\,\cos\theta
\qquad ,\qquad
a_y\cosh(\frac{d}{2\lambda})=-h_0\,\sin\theta
\,.
\end{equation}
The perturbation satisfies the inhomogeneous equation,
\begin{equation}
\frac{d^2\alpha_i}{d\zeta^2} - \alpha_i
= -L_i^2
\,\qquad i=x,y
\,.
\end{equation}
The solutions are obtained by writing $\alpha(\zeta)= {\cal
H}(\zeta)\phi(\zeta)$, where ${\cal H}(\zeta)$ is a solution to the
homogeneous equation. The function $\phi(\zeta)$ then satisfies,
\begin{equation}\label{phi}
\phi'' + \frac{2{\cal H}'}{{\cal H}}\,\phi' = -L^2/{\cal H}
\,.
\end{equation}
The first-order equation for $\phi'$ is obtained by multiplying
eq.(\ref{phi}) by the integrating factor ${\cal H}^2$,
\begin{equation}
\phi'(\zeta)=\frac{-1}{{\cal H}(\zeta)^2}\int_0^{\zeta}\,L^2(x)\,{\cal H}(x) dx
\,.
\end{equation}
The boundary conditions are satisfied at both interfaces by choosing the
homogeneous solution to be, ${\cal H} = \sinh(\zeta)$, in which case
\begin{equation}
\phi'(\zeta)=\frac{-a^2}{{\cal H}(\zeta)^2}\int_0^{\zeta}\sinh^3(x)\,dx
=-\frac{a^2}{3}\left[\cosh(\zeta)-{\rm sech}^2(\zeta/2)\right]
\,.
\end{equation}
One additional integration yields
\begin{equation}
\phi(\zeta)-\phi_0 = -\frac{a^2}{3}\left[\sinh(\zeta)-2\tanh(\zeta/2)\right]
\,,
\end{equation}
where $\phi_0$ is also chosen such that the full nonlinear solution
satisfies the boundary conditions. The resulting
solutions for the magnetic field are
\begin{equation}
b_x/H = -\frac{1}{h_0}\left(\frac{du_y}{d\zeta}\right)
=+\sin\theta\left[\beta(\zeta) + h_0\sin\theta\,\Phi(\zeta)\right]
\,,
\end{equation}
\begin{equation}
b_y/H = +\frac{1}{h_0}\left(\frac{du_x}{d\zeta}\right)
=-\cos\theta\left[\beta(\zeta) + h_0\cos\theta\,\Phi(\zeta)\right]
\,,
\end{equation}
\begin{equation}
\beta(\zeta)=\cosh(\zeta)/\cosh(\frac{d}{2\lambda})
\,,
\end{equation}
\begin{equation}
\Phi(\zeta)=\frac{8}{3}{\rm sech}^3(\frac{d}{2\lambda})\times
\left\{\cosh(\zeta)\sinh^3(\frac{d}{4\lambda})\cosh(\frac{d}{4\lambda})
- \cosh(\frac{d}{2\lambda})\sinh^3(\frac{\zeta}{2})\cosh(\frac{\zeta}{2})
\right\}
\,.
\end{equation}
Note that the nonlinear correction is largest at a distance of order
$\lambda/2$ from the interface.

\end{document}